\documentclass[10pt,a4paper]{article}

\newcommand\nc{\newcommand*} \nc\longnc{\newcommand}
\longnc\VOMIT[1]{#1}  \longnc\OMIT[1]{}

\usepackage[T1]{fontenc}
\usepackage[latin2]{inputenc}  
\usepackage{lmodern}

\usepackage{amsmath,amssymb}

\usepackage{graphicx}
\usepackage{float}

\nc\Section[1]{\section{#1}\setcounter{equation}{0}}

\def\ig[#1]#2{\includegraphics[#1]{Figs-arxiv/#2}}

\nc\rf[1]{Figure~\ref{#1}}  \nc\rff[1]{Figures~\ref{#1}}

\nc\xfffigg{\textwidth}  \nc\yfffigg{\vspace{0ex} \\}
\nc\fffigg[9]{
 \begin{figure}[H]  \centering
  \begin{minipage}[b]{#1\textwidth}
   \ig[width=\xfffigg]{#2}  \yfffigg
   \ig[width=\xfffigg]{#3}  \yfffigg
   \ig[width=\xfffigg]{#4}
  \end{minipage}
  \hfill
  \begin{minipage}[b]{#1\textwidth}
   \ig[width=\xfffigg]{#5}  \yfffigg
   \ig[width=\xfffigg]{#6}  \yfffigg
   \ig[width=\xfffigg]{#7}
  \end{minipage}
  \caption{#8}  \label{#9}
 \end{figure}
}
\nc\UpDown{\textsl{Up to down:} slow, medium and fast changes in the
  boundary condition.}
\nc\LeftRight[2]{\textsl{Left:} The coefficient function#1 belonging to
  stress. \textsl{Right:} The coefficient function#2 belonging to
  stress-dimensioned strain. \UpDown}

\nc\yfffig{\vspace{3ex} \\}
\nc\fffig[6]{
 \begin{figure}[H]  \centering
   \ig[width=#1\textwidth]{#2}  \yfffig
   \ig[width=#1\textwidth]{#3}  \yfffig
   \ig[width=#1\textwidth]{#4}
  \caption{#5}  \label{#6}
 \end{figure}
}

\nc\yfffiG{0.24\textheight}  \nc\YfffiG{\\ \vspace{0.03\textheight}}
\nc\Fbox[1]{\fbox{\rule[-.02\textheight]{0em}{.28\textheight}#1}}
\nc\fffiG[5]{
 \begin{figure}[H]  \centering
  \Fbox{\ig[height=\yfffiG]{#1}}  \YfffiG
  \Fbox{\ig[height=\yfffiG]{#2}}  \YfffiG
  \Fbox{\ig[height=\yfffiG]{#3}}
  \caption{#4}  \label{#5}
 \end{figure}
}

\nc\cf{cf.\ }  \nc\wrt{w.r.t.\ }
\nc\lat\emph  
\nc\ie{\lat{i.e.,\ }}  \nc\etal{\lat{et al.\ }}   \nc\etc{\lat{etc.\ }}
\nc\eg{\lat{e.g.,\ }}  \nc\insitu{\lat{in situ}}  \nc\QED{\lat{Q.E.D.}}
\nc\MerKoz{M\'er\-n\"ok\-ge\-o\-l\'o\-gia--K\H o\-zet\-me\-cha\-ni\-ka}

\nc\mathBf[1]{{\mathchoice{\hbox{\boldmath{$#1$}}}{\hbox{\boldmath{$#1$}}}
  {\hbox{\boldmath{$\scriptstyle #1$}}}          
  {\hbox{\boldmath{$\scriptscriptstyle #1$}}}}}  

\def\re#1{(\ref{#1})}  
\nc\m[1]{$           #1        $}  
\nc\mm[1]{$       \, #1 \,     $}  
\nc\mmm[1]{$    \,\, #1 \,\,   $}  
\nc\mmmm[1]{$ \,\,\, #1 \,\,\, $}  

\nc\bi\relax                      
\nc\bit{    \mskip1mu}  \nc\biT{    \mskip-1mu}  
\nc\bitt{   \mskip2mu}  \nc\biTT{   \mskip-2mu}  
\nc\bittt{  \mskip3mu}  \nc\biTTT{  \mskip-3mu}  
\nc\bitttt{ \mskip4mu}  \nc\biTTTT{ \mskip-4mu}  
\nc\bittttt{\mskip5mu}  \nc\biTTTTT{\mskip-5mu}  

\nc\0[2]{\ifcase#1{#2}\or\lt(#2\rt)\or\lt[{#2}\rt]\or\lt\{{#2}\rt\}\or
 \mathord<{#2}\mathord>\or\lt\langle{#2}\rt\rangle\or\lt\lvert{#2}\rt
 \rvert\or\lt\lVert{#2}\rt\rVert\fi}
\nc\1[2]{\ifcase#1{#2}\or(#2)\or[#2]\or\{#2\}\or\mathord<{#2}\mathord
 >\or\langle{#2}\rangle\or\lvert{#2}\rvert\or\lVert{#2}\rVert\fi}
\nc\2[2]{\ifcase#1{#2}\or\big(#2\big)\or\big[#2\big]\or\big
 \{#2\big\}\or\big<#2\big>\or\big\langle#2\big\rangle\or\big
 \lvert#2\big\rvert\or\big\lVert#2\big\rVert\fi}
\nc\3[2]{\ifcase#1{#2}\or\Big(#2\Big)\or\Big[#2\Big]\or\Big\{#2\Big
 \}\or\Big<#2\Big>\or\Big\langle#2\Big\rangle\or\Big\lvert#2\Big
 \rvert\or\Big\lVert#2\Big\rVert\fi}
\nc\4[2]{\ifcase#1{#2}\or\bigg(#2\bigg)\or\bigg[#2\bigg]\or\bigg
 \{#2\bigg\}\or\bigg<#2\bigg>\or\bigg\langle#2\bigg\rangle\or\bigg
 \lvert#2\bigg\rvert\or\bigg\lVert#2\bigg\rVert\fi}
\nc\5[2]{\ifcase#1{#2}\or\Bigg(#2\Bigg)\or\Bigg[#2\Bigg]\or\Bigg
 \{#2\Bigg\}\or\Bigg<#2\Bigg>\or\Bigg\langle#2\Bigg\rangle\or\Bigg 
 \lvert#2\Bigg\rvert\or\Bigg\lVert#2\Bigg\rVert\fi}
\nc\9[2]{\ifcase#1{#2}\or\left(#2\right)\or\left[#2\right]\or\left
 \{#2\right\}\or\left\langle{#2}\right\rangle\or\left\langle{#2}\right
 \rangle\or\left\lvert{#2}\right\rvert\or\left\lVert{#2}\right\rVert\fi}
\nc\lt{\mathopen{}\mathclose\bgroup\left} \nc\rt{\aftergroup\egroup\right}
\nc\7[2]{\mskip.5mu \1#1{#2}}
\nc\8[2]{\1#1{#2}\bitt}

\nc\f\frac
\nc\F[2]{#1/#2}
\nc\tr{\operatorname{tr}}

\nc\tensor\mathbf  \nc\Tensor\mathBf

\def\lta#1{{\overset{{\scriptscriptstyle \leftarrow}}{#1}}}
\def\rta#1{{\overset{{\scriptscriptstyle \rightarrow}}{#1}}}
\def\nablal{\lta{\nabla}}      \def\nablar{\rta{\nabla}}

\nc\dev[1]{#1^{\rm d}}         \nc\sph[1]{#1^{\rm s}}
\nc\devv[1]{#1\bit^{\rm d}}    \nc\sphh[1]{#1\bit^{\rm s}}
\nc\devs[1]{\dev{#1}_\qsig}    \nc\sphs[1]{\sph{#1}_\qsig}
\nc\deve[1]{\dev{#1}_\qeps}    \nc\sphe[1]{\sph{#1}_\qeps}
\nc\el[1]{#1_{\rm el}}
\nc\elt[1]{#1_{{\rm el},\qlam}}
\nc\rhe[1]{#1_{\rm rheol}}
\nc\Cauchy{{^{\rm Cauchy}}}

\nc\qdt{\f{\pd}{\pd t}}    \nc\qddt{\f{\pd^2}{\pd^2 t}}
\nc\qDt{\f{\dd}{\dd t}}    \nc\qDDt{\f{\dd^2}{\dd^2 t}}

\nc\dd{\mathrm{d}}
\nc\pd\partial

\nc\ident{\tensor1}  \nc\zero{\tensor0}

\nc\oE{\mathcal{E}}  \nc\oS{\mathcal{S}}  \nc\oZ{\mathcal{Z}}

\nc\qE{E}  \nc\qEE{{\hat{\qE}}}  \nc\qEEE{{\hat{\qEE}}}
\nc\qEd{\dev\qE}           \nc\qEs{\sph\qE}
\nc\qEds{\qEd_\qsig}       \nc\qEss{\qEs_\qsig}
\nc\qEde{\qEd_\qeps}       \nc\qEse{\qEs_\qeps}

\nc\qrho{\varrho}
\nc\qg{\tensor g}
\nc\qeta{\eta}  \nc\qetas{\qeta_\qsig}  \nc\qetae{\qeta_\qeps}

\nc\qr{\tensor{r}}
\nc\qrt{\tensor{\tilde{r}}}
\nc\RI{R_1}  \nc\RO{R_2}
\nc\qtet{\vartheta}

\nc\qeps{\varepsilon}      \nc\qsig{\sigma}           \nc\qzet{\zeta}
\nc\qqeps{\Tensor{\qeps}}  \nc\qqsig{\Tensor{\qsig}}  \nc\qqzet{\Tensor{\qzet}}
\nc\heps{\hat{\qeps}}      \nc\hsig{\hat{\qsig}}
\nc\hheps{\hat{\qqeps}}    \nc\hhsig{\hat{\qqsig}}    \nc\hhzet{\hat{\qqzet}}
\nc\beps{\bar{\qeps}}      \nc\bsig{\bar{\qsig}}
\nc\bbeps{\bar{\qqeps}}    \nc\bbsig{\bar{\qqsig}}    \nc\bbzet{\bar{\qqzet}}
\nc\deps{\dot{\qeps}}    \nc\Deps{\dot{\qqeps}}
\nc\ddeps{\ddot{\qeps}}  \nc\DDeps{\ddot{\qqeps}}
\nc\qqu{\tensor{u}}
\nc\qOm{\Tensor{\Omega}}
\nc\qs{\tensor s}

\nc\qa{a}  \nc\qb{b}  \nc\qc{c}  \nc\qd{d}
\nc\qA{A}  \nc\qB{B}  \nc\qC{C}  \nc\qD{D}
\nc\qalpha{\Tensor{\alpha}}  \nc\qbeta{\Tensor{\beta}}
\nc\qgamma{\Tensor{\gamma}}  \nc\qdelta{\Tensor{\delta}}
\nc\qPsid{\dev{\Psi}}  \nc\qPsis{\sph{\Psi}}
\nc\qPhid{\dev{\Phi}}  \nc\qPhis{\sph{\Phi}}
\nc\qphi{\varphi}
\nc\qpsi{\psi}

\nc\qlam{\lambda}
\nc\qkap{\kappa}

\begin{document}


\title{Analytical solution method for rheological problems of solids}
\author{Tamás Fülöp\thanks{Corresponding author,
\texttt{fulop@energia.bme.hu}.},  Mátyás Szücs
 \\
 \\
Department of Energy Engineering,
 \\
Faculty of Mechanical Engineering,
 \\
Budapest University of Technology and Economics,
 \\
Budapest H-1521, Hungary;
 \\
Montavid
Research Group, Budapest, Hungary
}
 
 \maketitle

 \begin{abstract}
In classical continuum theory, Volterra's principle \cite{gromov,
rabotnov} is a long-known method to solve linear rheological
(viscoelastic) problems derived from the corresponding elastic ones.
Here, we introduce and present another approach that is simpler to apply
(no operator inverse is required to compute but only linear ordinary
differential equations to solve). Our method starts with the known
elastic solution, replaces the elasticity coefficients with time
dependent functions, derives differential equations on them, and
determines the solution corresponding to the initial conditions. We
present several examples solved via this new method, like tunnels and
spherical hollows opened in various initial stress states, and 
pressurizing of thick-walled tubes and spherical tanks. These examples
are useful for applications and, in parallel, are suitable for testing
and validating numerical methods of various kinds.
 \end{abstract}

\noindent\textbf{Keywords:} solids, elasticity, rheology,
viscoelasticity, analytical solution, Volterra's principle, displacement
field

\Section{Motivation and introduction}

A large variety of solid materials---like plastics, rocks, asphalt,
biomaterials \etc---possess viscoelastic/rheological characteristics.
Correspondingly, one can observe some kind of delayed and damped elastic
behaviour.
 
Rheological behaviour of solid media is well-known, \eg in civil
engineering and in mine industry.
A hollow opened in an underground stone block often takes its eventual
shape only years after the drilling. The
diagrams in \rf{pic1} demonstrate this.

 \begin{figure}[H]  \centering
   \ig[width=0.475\textwidth]{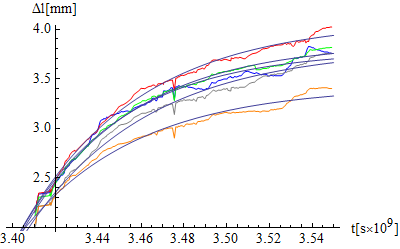}
  \hfill
   \ig[width=0.475\textwidth]{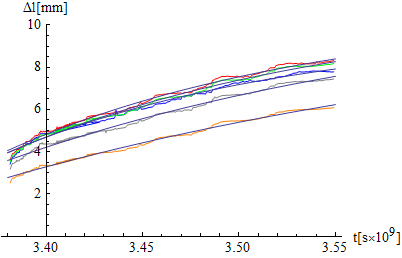}
\caption{%
Measured (and fitted) exponential-like displacement history of tunnel
walls at the National Radioactive Waste Repository, Bátaapáti, Hungary,
with characteristic times of 3--10 years \cite{kovacs} (different
colours mean different directions at a given cross-section of the
tunnel).}
  \label{pic1}
 \end{figure}

Accordingly, rheological behaviour must be taken into account when
designing technical devices and facilities. Nevertheless, this means not
only disadvantages and problems but also benefits. For example, one can
rely on its effect of damping and absorbing vibrations. Many
biology-originated objects and protheses used in medical technology also
show rheological properties. For instance, rheological behaviour of knee
ligaments is apparent. Similarly, a freshly opened underground tunnel
needs initially only a temporary---relatively weak---support, and the
eventual support is enough to be established only later, when most of
displacements have already been occured, allowing thus a much cheaper
eventual support.
 
However, all these need reliable calculations. Nowadays, the most often
applied methods are the discretization based numerical methods, which
face at problems. Solutions obtained by such methods may considerably
depend on the resolution of the applied discretization. Furthermore, for
complex three-dimensional problems calculation times are large.
Therefore, analytical solution of a simplified version of the problem
may provide a reasonable first approximation and give useful insight.
Analytical solutions can also be utilized for validating numerical
methods.
 
Here, we introduce and present an exact analytical method for solving
linear rheological problems of solids. The approach is based on the
corresponding elastic solutions assumed to be already known: the
elasticity coefficients are replaced with time dependent functions,
which are determined from the rheological equations. The method has been
born in a conceptually simple form, with limited range of applicability,
and has been enhanced and generalized subsequently in two further steps.
Here, we present these three stages in order of increasing generality.
For each stage, we show several examples, which not only illustrate the
method but also demonstrate its power and limitations.

\Section{Elasticity and rheology}  \label{elrheol}

We are going to treat purely mechanical problems of homogeneous and
isotropic continuous media, and our aim is to determine the displacement
field \m{\qqu}, the strain field \m{\qqeps} and the stress field
\m{\qqsig} \1 1 {where both \m { \qqeps } and \m { \qqsig } are
symmetric tensors}. We wish to work in the force equilibrial
approximation, \ie when acceleration is neglected:\footnote{Accordingly,
wave phenomena are omitted from our scope. Nevertheless, nontrivial time
dependent processes will emerge, as an interplay of the time dependent
boundary conditions and the rheological material model.}
 \begin{align}  \label{imp}
\qqsig\cdot\nablal&=-\qrho\qg,
 \end{align}
where \m { \qrho } is mass density and \m{\qrho\qg} is volumetric force
density \1 1 {assumed to be time independent}, and \m{\nablal} and
\m{\nablar} are the nabla operators acting to the left and to the right,
respectively \1 1 {reflecting proper tensorial order, see also below}.

Concerning \m { \qqeps }, we stay in the small-strain approximation, 
which then imposes the geometric compatibility equation in the form
 \begin{align}  \label{comp}
\nablar\times\qqeps\times\nablal&=\zero.
 \end{align}
According to mathematics, to a symmetric tensor field \m { \qqeps } with
property \re{comp}, there exists a vector field \m { \qqu\Cauchy
}---called hereafter Cauchy vector potential
\cite{fulop_rug_reol}---from which \m { \qqeps } can be obtained as
 \begin{align}
\qqeps=\left(\qqu\Cauchy\otimes\nablal\right)^{\rm Sym} ,
 \end{align}
where \m{^{\rm Sym}} denotes the symmetric part of a tensor. The Cauchy
vector potential is not unique for a given \m { \qqeps }, and all Cauchy
vector potentials can be derived from the strain field according to
Ces\`aro's formula \cite{beda_kozak_verhas_kont_mech},
 \begin{align}  \label{cesaro}
\qqu\Cauchy\01{t,\qr}=\qqu_0\01{t}+\qOm\81{t}\01{\qr-\qr_0}+\int_{\qr_0}^\qr
\93{\qqeps\01{t,\qrt}+2\92{\qqeps\01{t,\qrt}\otimes\nablal}^{{\rm
A}_{1,3}}\01{\qr-\qrt}}\dd \qrt
 \end{align}
with \m{^{{\rm A}_{1,3}}} denoting antisymmetrization in the first and
third indices, where the position vector \m{\qr_0}, the path of
integration, the vector function \m{\qqu_0\01{t}} and the antisymmetric
tensor function \m{\qOm\01{t}} are each arbitrary. The displacement
field \m { \qqu } is one of these Cauchy vector potentials so when we
wish to reconstruct \m{\qqu} from the strain field then we need to fix
these uncertainties using symmetry arguments and other physically
plausible considerations.

In case of linear elasticity (for a homogeneous and isotropic medium),
connection between stress and strain is provided by Hooke's law,
 \begin{align}  \label{hooke}
\qqsig=\dev{\qqsig}+\sph{\qqsig},
 \qquad\qquad
\dev{\qqsig}=\qEd\dev{\qqeps},
 \qquad\quad
\sph{\qqsig}=\qEs\sph{\qqeps}
 \end{align}
in the deviatoric--spherical separation, where
\mm{\sph{\qqsig}=\f{1}{3}\81{\tr\qqsig}\ident} denotes the spherical
part---which is proportional to the identity tensor \m{\ident}---, while
\mm{\dev{\qqsig}=\qqsig-\sph{\qqsig}} is the deviatoric (traceless)
part; furthermore, \m{\qEd=2G} is the deviatoric elasticity
coefficient and \m{\qEs=3K} is the spherical one\footnote{\m{G} is
the shear modulus and \m{K} is the bulk modulus.}.

For linear rheological models of solids, one can generalize Hooke's law
by replacing the elasticity coefficients with polynomials of the time
derivative operator. Namely,
 \begin{align}  \label{reol}
 \dev{\oS}\dev{\qqsig}=\dev{\oE}\dev{\qqeps},
\qquad\qquad
 \sph{\oS}\sph{\qqsig}=\sph{\oE}\sph{\qqeps},
 \end{align}
where the stress related operators \m{\dev{\oS}}, \m{\sph{\oS}} and the
strain related ones \m{\dev{\oE}}, \m{\sph{\oE}} are
 \begin{align}  \label{op1}
\dev{\oS} & =1+\dev{\tau}_1\f{\pd}{\pd t}+\dev{\tau}_2\f{\pd^2}{\pd
t^2}+\cdots,  &  \dev{\oE} & =\qEd_0+\qEd_1\f{\pd}{\pd
t}+\qEd_2\f{\pd^2}{\pd t^2}+\cdots,
 \\  \label{op2}
\sph{\oS} & =1+\sph{\tau}_1\f{\pd}{\pd t}+\sph{\tau}_2\f{\pd^2}{\pd
t^2}+\cdots,  &  \sph{\oE} & =\qEs_0+\qEs_1\f{\pd}{\pd
t}+\qEs_2\f{\pd^2}{\pd t^2}+\cdots
 \end{align}
with constant coefficients \m { \dev{\tau}_i, \sph{\tau}_j, \qEd_k,
\qEs_l }. In our applications, we concentrate on the
Kluitenberg--Verh\'as model family \cite{asszonyi_fulop_van}
 \begin{align}
\dev{\qqsig}+\dev{\tau}\devv{\dot\qqsig}&=\qEd_0\dev{\qqeps}+\qEd_1\dev{\Deps}+\qEd_2\dev{\DDeps},
 &
\sph{\qqsig}+\sph{\tau}\sphh{\dot\qqsig}&=\qEs_0\sph{\qqeps}+\qEs_1\sph{\Deps}+\qEs_2\sph{\DDeps},
 \end{align}
which is important from both theoretical \cite{asszonyi_fulop_van} and
experimental \cite{lin_etal,matsuki1,matsuki2,asszonyi_csatar_fulop}
aspects, and covers various classic rheological models as special cases;
hereafter, overdot abbreviates partial time derivative\footnote{The
small-strain assumption allows to approximate the substantial time
derivative with the partial time derivative.}.
 
In the case of elasticity, equations \re{imp}, \re{comp} and \re{hooke}
form the system of equations to be solved. Together with appropriate
boundary conditions imposed at the boundaries of the spatial domain
considered, the solution exists and is unique, however, to obtain this
solution is not necessarily simple, since \re{hooke} poses
\emph{separate} conditions for the deviatoric and spherical parts,
while \re{imp}, \re{comp} and the boundary conditions prescribe
requirements for the \emph{sum} of the deviatoric and spherical parts.
 
When we deal with the above-described rheological generalization of the
problem then, in addition to \re{imp}, \re{comp}, \re{reol} and the
boundary conditions \1 1 {which may be time dependent in general},
initial conditions are also required to ensure uniqueness of the
solution, since the constitutive equations contain time derivatives [see
\re{op1} and \re{op2}]. In the rheological case, all fields are
functions of both time and space, and all equations and boundary
conditions have to be satisfied for all time instants, which raises an
even more complicated task than for the elastic counterpart. It would
considerably simplify the situation if one could utilize the known space
dependence of the corresponding elastic problem, leaving only time
dependence to address.
 
Volterra's principle \cite{gromov, rabotnov} provides such an
opportunity, according to which principle the constants \m{\qEd},
\m{\qEs} in the elastic solution are to be replaced with the
rheological operators \m{\dev{\oS}}, \m{\sph{\oS}}, \m{\dev{\oE}},
\m{\sph{\oE}}, and solving the resulting temporal equations leads to the
solution of the rheological problem. However, in some cases the
application of Volterra's principle is difficult, \eg in cases of time
dependent stress boundary conditions, in addition to the fact that, in
Volterra's approach, typically one also has to invert operators.

Below, we present another route, which is also motivated by Volterra's
idea to treat the space dependence aspect of the problem via utilizing
the known solution of the corresponding elastic problem, but is
technically easier to follow since only a set of ordinary differential
equations is to be solved for the time dependence aspect.

\Section{The analytical solution method for the rheological problem}
\label{anmeth}

As the first step, let us separate the effect of the force density by
subtracting some such time independent fields \m{\bbsig} and \m{\bbeps}
---henceforth: primary fields--- that satisfy the equations
 \begin{align}
\bbsig\cdot\nablal & = -\qrho\qg,
 \\
\nablar\times\bbeps\times\nablal & = \zero ,
 \\
\dev{\bbsig} = \qEd_0 \dev{\bbeps},
 \quad & \mathrel{\qquad}
\sph{\bbsig} = \qEs_0 \sph{\bbeps}
 \end{align}
[note that, for time independent stress and strain fields, \re{reol}
gets simplified to Hooke's law with \mm { \qEd = \qEd_0 }, \mm {
\qEs = \qEs_0 }]. Then the difference fields---henceforth:
complementary fields---
 \begin{align}
 &&
\hhsig&:=\qqsig-\bbsig,
 &
\hheps&:=\qqeps-\bbeps
 &&
 \end{align} 
satisfy the homogeneous equations
 \begin{align}  \label{impp}
\hhsig\cdot\nablal & = \zero ,
 \\  \label{compp}
\nablar\times\hheps\times\nablal & = \zero ,
 \\  \label{reoll}
\dev{\oS}\dev{\hhsig} = \dev{\oE}\dev{\hheps},
\quad & \mathrel{\qquad}
\sph{\oS}\sph{\hhsig} = \sph{\oE}\sph{\hheps} .
 \end{align}

Naturally, in general, this transformation modifies the boundary
conditions, which is to be taken into account during the calculations.

If the spatial domain filled with the medium has more than one boundary
\1 1 {more than one connected boundary surface} then the problem can be
divided into subproblems in which only one boundary condition is nonzero
\1 1 {the boundary condition is nonzero only on one connected boundary
surface}. Henceforth, we always analyze one such subproblem. Thanks to
linearity of all equations involved, the sum of such subsolutions
provides solution for a whole problem.

In this work, we consider problems with time dependent boundary
conditions that are prescribed for \emph{stress}, not for
\emph{displacement}. The applications discussed here will all be
related to such boundary conditions.

Time dependence of boundary conditions will be allowed with the
limitation that time dependence must mean a space independent rescaling
of the boundary condition---like gradual loading of a surface where
loading may be space dependent but the ratio of normal stress values at
two different boundary points is time
independent.\footnote{Illustratively speaking, the boundary condition
must realize homogeneous amplification/tuning along the boundary.} In
notation, time dependence of the boundary condition is of the form of a
time dependent multiplier \m{\qlam\01{t}}. This \m{\qlam\01{t}} can be
quite arbitrary, the only restriction being that it be sufficiently many
times differentiable. For gradual switching on, like when modelling
drilling, this factor can be chosen as

 \noindent
 \unitlength=.107em 
 \hskip-.4em
 \parbox[b]{11em}{%
  \raisebox{-.4ex}{\mbox{%
    \begin{picture}(100,58)(0,0) 
    \put(9,15){\vector(1,0){80}}
    \put(92,15){\makebox(0,0)[l]{\scriptsize{\m{t}}}}
    \put(53,39){{\scriptsize{\m{\qlam(t)}}}}
    \put(15,11){\vector(0,1){40}}
    \put(44,12){\line(0,1){6}}
    \put(44.3,11){\makebox(0,0)[t]{\scriptsize{\m{t_1}}}}
    \put(75,12){\line(0,1){6}}
    \put(75.3,11){\makebox(0,0)[t]{\scriptsize{\m{t_2}}}}
    \put(12,37){\line(1,0){6}}
    \put(9,37){\makebox(0,0){\scriptsize{\m{1}}}}
    \thicklines
    \put(15,15){\line(1,0){29}} \qbezier(62,28)(59,15)(44,15)
    \put(74,37){\line(1,0){14}} \qbezier(62,28)(65,37)(74,37)
    \end{picture}%
 }}}%
 \hfill
 \parbox[b]{23.8em}{
 \begin{align}  \label{lambda}
\qlam(t) = \left\{  \begin{matrix}
0 & {\rm if} & t \le t_1,
 \\
1 & {\rm if} & t \ge t_2,
 \\
{\text{smooth in between.}}
 &&
 \end{matrix}  \right.
 \end{align}
 }
 \linebreak
Corresponding to such a time dependent homogeneous rescaling of the
boundary condition, the solution of the \emph{elastic} problem also
gets rescaled---space independently rescaled---by the factor \mm {
\qlam(t) }.

Another remarkable property of the elastic solution \1 1 {at any fixed
\m { t }} is that, based on dimensional reasoning, dependence of the
stress solution on \m{\qEd} and \m{\qEs} must be such that stress
depends only on the dimensionless ratio
 \begin{align}  \label{eta}
\qeta:=\f{\qEd}{\qEs} .
 \end{align}
In other words, it depends only on the Poisson's ratio
 \begin{align}  \label{nu}
\nu = \f{1-\qeta}{2+\qeta}
 \end{align}
This property is, naturally, apparently visible in the examples
considered below. It is to be emphasized that, although usually one
focuses only on the
{space dependence} of an elastic solution, for the solution methods
described here, dependence on the elasticity coefficients \m{\qEd},
\m{\qEs} will also be of central importance.

\subsection{First approach: Method of elasticity constants made time
dependent}
\label{method1}

In the first---the simplest---version of our approach, we adapt
Volterra's principle in the form that the rheological solution is
obtained from the elastic one by replacing the elasticity
coefficients \m{\qEd}, \m{\qEs} with time dependent functions
\1 1 {rather than with operators, as in Volterra's methodology}.

At any instant \m { t }, for any \m{\qEd\01{t}}, \m{\qEs\01{t}}
the solution is a valid elastic solution (for the current boundary
condition) so \re{impp} and \re{compp}---as well as the boundary
condition---are satisfied. 

Apparently,  the spatial equations---\re{impp}, \re{compp} and the
stress boundary condition---will be satisfied at any time instant, with
the actual values \m{\qEd\01{t}}, \m{\qEs\01{t}}. The remaining
equations---namely, the rheological ones \re{reoll}---will yield
ordinary differential equations on \m{\qEd\01{t}} and
\m{\qEs\01{t}}.
 
One can observe that one part of the spatial conditions, \re{impp} and
the stress boundary condition, refers only to the stress field while the
remaining spatial condition, \re{compp}, refers only to the strain
field. Accordingly, one is allowed to detune the elastic solution for
stress from the elastic solution for strain: the elastic solution for
stress can be utilized with some \m{\devs{\qE}\01{t}}, \m{\sphs{\qE}\01{t}}
while the elastic solution for strain can contain some separate
\m{\deve{\qE}\01{t}}, \m{\sphe{\qE}\01{t}}. Notably, the only condition
forbidding this detuning would be \re{hooke} but for rheology it is
replaced with \re{reoll} so it is not excluded that some consistent
solution can be found.

As can be seen in the examples below, consistent solutions are indeed
possible, either with this detuning or even without it.

\subsubsection{Cylindrical bore (tunnel) opened in infinite, homogeneous
and isotropic stress field}  \label{izotrop}

In an infinite, homogeneous and isotropic stress field \m { \bbsig }, we
open an infinite cylindrical bore with radius \m { R } (see
\rf{outlineizo}). In cylindrical coordinates, the boundary
conditions specifying the elastic solution for the completely open bore
are
 \begin{align}  \label{izobound}
\qsig_{rr}(R,\qphi,z)=0,
\qquad\qquad\qquad
\lim_{r \to \infty}\qqsig(r,\qphi,z)=\bbsig,
 \end{align}
which are rewritten for the complementary field as
 \begin{align}
\hsig_{rr}(R,\qphi,z)=-\bsig_{rr},
 \qquad\qquad\quad
\lim_{r \to \infty}\hhsig(r,\qphi,z)=\zero.
 \end{align}

 \begin{figure}[H]  \centering
   \ig[width=0.4\textwidth]{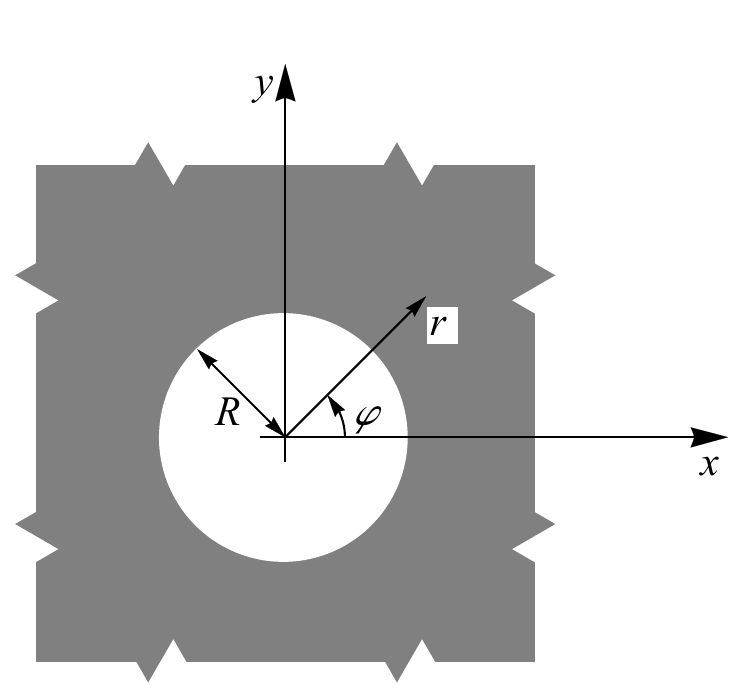}
\caption{Outline and notations for the cylindrical bore in infinite,
homogeneous and isotropic stress field.}
  \label{outlineizo}
 \end{figure}

The solution of the elastic problem, for this completely opened bore, is
 \begin{align}
\el\hhsig(\qr)=\bsig_{rr}\f{R^2}{r^2}\begin{pmatrix}-1&0&0\\0&1&0\\0&0&0\end{pmatrix}.
 \end{align}
Notice that the spherical part of this tensor is zero,
\m{\sph{\hhsig}=\zero}, therefore, \m{\hhsig=\dev{\hhsig}}, from which the
strain tensor is
 \begin{align}
\el\hheps(\qr)=\f{1}{\dev{\qE}}\dev{\el\hhsig}(\qr)=\dev{\el\hheps}(\qr).
 \end{align}
 
Now let us consider the problem with time dependent boundary condition,
\eg with a \m { \qlam(t) } of the form \re{lambda} via which we can
model the drilling process. The corresponding elastic solution is
nothing but the previous one rescaled by \m { \qlam(t) }:
 \begin{align}
\elt\hhsig^{} \71{t, \qr} & = \dev{\elt\hhsig} \71{t, \qr} =
\qlam(t) \cdot \dev{\el\hhsig}(\qr),
 \\
\elt\hheps^{} \71{t, \qr} & = \dev{\elt\hheps} \71{t, \qr} =
\f{\qlam(t)}{\dev{\qE}} \cdot \dev{\el\hhsig}(\qr) .
 \end{align}
For a rheological problem \re{reoll}, the above-described method of time
dependent elastic coefficients says to substitute the only elasticity
coefficient \m{\dev{\qE}} present in the previous equations for an
unknown time dependent function \m { \dev{\qE}(t) }:
 \begin{align}
\rhe\hhsig \71{t, \qr} & = \qlam(t)\cdot\dev{\el\hhsig}(\qr),
 \\
\rhe\hheps \71{t, \qr} & = \f{\qlam(t)}{\dev{\qE}(t)}\cdot\dev{\el\hhsig}(\qr).
 \end{align}
Obviously,  the spatial conditions are satisfied at any time \m { t }.
The only thing left to do is to impose \re{reoll}.
The spherical equation is trivially fulfilled, while
 \begin{align}  \label{izoalagutreol}
\dev{\oS}\qlam(t) &= \dev{\oE}\f{\qlam(t)}{\dev{\qE}(t)}
 \end{align}
is generated for the deviatoric part. We have to solve this equation
with such initial conditions that the complementary stress and strain
fields before the drilling (before the time dependent change in the
boundary condition) are zero for an extended time interval---implying
that all time derivatives of the fields are also zero. Then the solution
of \re{izoalagutreol} is unique so the method has reached the goal.

Comparing the outcome found here with the one known in the literature
\cite{izoalaguteng}, obtained via another approach, we find that the two
results are in complete agreement.

For concrete rheological models, \ie for concrete rheological operators
\re{op1}--\re{op2}, solutions will be presented and plotted in
Section~\ref{concrete}.\footnote{The same holds for all the subsequent
problems, too: For plots obtained for concrete rheological models, see
Section~\ref{concrete}.}

\subsubsection{The rheological process of a spherical hollow in an
infinite, homogeneous and isotropic stress field}  \label{hollow}

The boundary conditions for a completely opened spherical hollow of
radius \m { R } opened in infinite, homogeneous and isotropic stress
field \m { \bbsig } (see \rf{outlinehollow}) are, in spherical
coordinates:
 \begin{align}
 \qsig_{rr}(R, \qtet, \qphi)=0,
 \qquad\qquad\qquad
 \lim_{r \to \infty} \qqsig(r, \qtet, \qphi)=\bbsig,
 \end{align}
rewritten for the complementary field as
 \begin{align}
 \hsig_{rr}(R, \qtet, \qphi)=-\bsig_{rr} ,
 \qquad\qquad\quad
 \lim_{r \to \infty} \hhsig(r, \qtet, \qphi)=\zero .
 \end{align}

 \begin{figure}[H]  \centering
   \ig[width=0.4\textwidth]{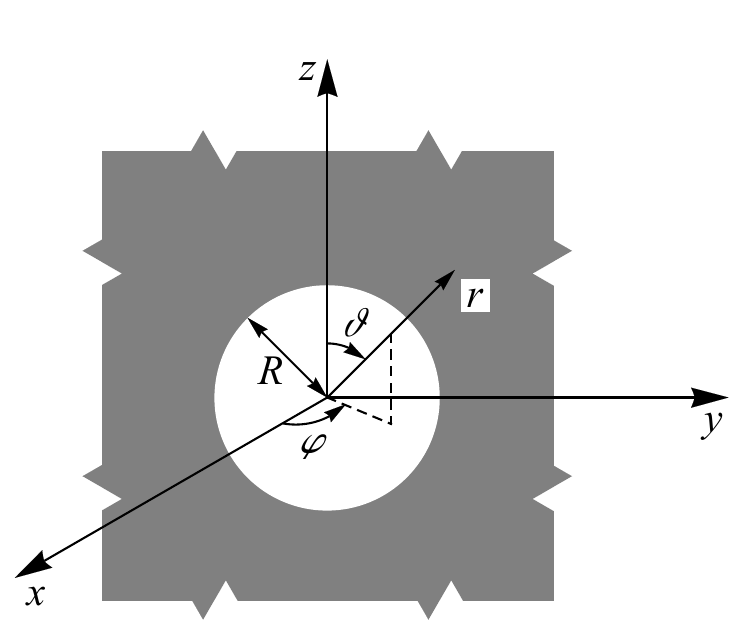}
\caption{Outline and notations for the spherical hollow in infinite,
homogeneous and isotropic stress field.}
  \label{outlinehollow}
 \end{figure}

The elastic stress solution is
 \begin{align}
\el\hhsig(\qr)=\bsig_{rr}\f{R^3}{r^3}\begin{pmatrix}-1&0&0\\ 0&\f{1}{2}&0\\ 0&0 &\f{1}{2}\end{pmatrix}.
 \end{align}
The spherical part of this tensor is zero again so
\m{\sph{\hhsig}=\zero}, \m{\dev{\hhsig}=\hhsig}, hence, the strain tensor
is
 \begin{align}
\el\hheps(\qr)=\f{1}{\dev{\qE}} \dev{\el\hhsig}(\qr)=\dev{\el\hheps}(\qr).
 \end{align}
Also analogously to the previous, cylindrical, case, the solution of the
elastic problem for the gradually opened hollow
is
 \begin{align}
\elt\hhsig^{} \71{t, \qr} & = \dev{\elt\hhsig} \71{t, \qr} =
\qlam(t) \cdot \dev{\el\hhsig}(\qr),
 \\
\elt\hheps^{} \71{t, \qr} & = \dev{\elt\hheps} \71{t, \qr} =
\f{\qlam(t)}{\dev{\qE}} \cdot \dev{\el\hhsig}(\qr).
 \end{align}

For rheology, we substitute the only elasticity coefficient
\m{\dev{\qE}} for an unknown time dependent function:
 \begin{align}
\rhe\hhsig \71{t, \qr} & = \qlam(t)\cdot\dev{\el\hhsig}(\qr),
 \\
\rhe\hheps \71{t, \qr} & = \f{\qlam(t)}{\dev{\qE}(t)}\cdot\dev{\el\hhsig}(\qr).
 \end{align}
When the rheological operators act on these functions then the spherical
equation is trivially fulfilled, while
 \begin{align}  \label{gomb}
\dev{\oS}\qlam(t) &= \dev{\oE}\f{\qlam(t)}{\dev{\qE}(t)}
 \end{align}
follows for the deviatoric part. With initial conditions as
in the previous example---zero initial history---the solution exists and
is unique.

Comparing \re{gomb} with \re{izoalagutreol} shows that, from the point
of view of our method, these two problems lead to the same rheological
equation (there are only spatial differences between the two problems).
The same similarity is the reason why we treat the two next problems,
pressurizing of a thick-walled tube and of a spherical tank, together.

\subsubsection{Pressurizing of a thick-walled tube and of a spherical
tank}  \label{tubetank}

Our next examples are the pressurizing of a thick-walled tube and of a
spherical tank from zero overpressure to overpressure \m{p_0}. Effect of
of possible initial pressure can be subtracted by the primary fields so
we formulate the problem directly for the complementary fields.
 
The boundary conditions for a thick-walled tube
with the inner and outer radia \m{\RI}, \m{\RO} at overpressure \m{p_0} are
 \begin{align}
\hsig_{rr}(\RI, \qtet, \qphi)=-p_0,
 \qquad\qquad\qquad
\hsig_{rr}(\RO, \qtet, \qphi)=0.
 \end{align}
The elastic stress solution in the wall is
 \begin{align}
\el\hhsig(\qr)=p_0\f{\RI^2}{\RO^2-\RI^2}\begin{pmatrix}1-\f{\RO^2}{r^2}&0&0 \\0&1+\f{\RO^2}{r^2}&0\\0&0&0\end{pmatrix},
 \end{align}
 which can separated into a deviatoric and a spherical part as
 \begin{align}
\dev{\el\hhsig}(\qr) & =
p_0\f{\RI^2}{\RO^2-\RI^2}\begin{pmatrix}\f{1}{3}-\f{\RO^2}{r^2}&0&0\\ 0&
\f{1}{3}+\f{\RO^2}{r^2}&0\\ 0& 0& -\f{2}{3}\end{pmatrix},
 &&&
\sph{\el\hhsig}(\qr) &=\f{2}{3}p_0\f{\RI^2}{\RO^2-\RI^2}\begin{pmatrix}1&0&0\\ 0& 1&0\\ 0&0 & 1\end{pmatrix}.
 \end{align}
Hence, the strain tensor is
 \begin{align}  \label{csostrain}
\el\hheps(\qr) = \f{1}{\dev{\qE}} \dev{\el\hhsig}(\qr) +
\f{1}{\sph{\qE}} \sph{\el\hhsig}(\qr).
 \end{align}
 
Similarly, the boundary conditions for a spherical tank
with inner and outer radia \m{\RI}, \m{\RO} are
 \begin{align}
\hsig_{rr}(\RI, \qtet, \qphi)=-p_0,
 \qquad\qquad\qquad
\hsig_{rr}(\RO, \qtet, \qphi)=0.
 \end{align}
 The elastic stress solution is
 \begin{align}
\el\hhsig(\qr) = p_0\f{\RI^3}{\RO^3-\RI^3}
 \begin{pmatrix}
1-\f{\RO^3}{r^3}&0&0\\ 0& 1+\f{\RO^3}{2r^3}&0\\ 0&0 & 1+\f{\RO^3}{2r^3}
 \end{pmatrix} ,
 \end{align}
the deviatoric and spherical parts of which are
 \begin{align}
\dev{\el\hhsig}(\qr)
&=p_0\f{\RI^3}{\RO^3-\RI^3}\begin{pmatrix}-\f{\RO^3}{r^3}&0&0\\ 0&
\f{\RO^3}{2r^3}&0\\ 0& 0& \f{\RO^3}{2r^3}\end{pmatrix},
 &
\sph{\el\hhsig}(\qr) &=p_0\f{\RI^3}{\RO^3-\RI^3}\begin{pmatrix}1&0&0\\ 0& 1&0\\ 0& 0& 1\end{pmatrix},
 \end{align}
and the corresponding strain tensor is
 \begin{align}  \label{uregstrain}
\el\hheps(\qr) = \f{1}{\dev{\qE}} \dev{\el\hhsig}(\qr) +
\f{1}{\sph{\qE}} \sph{\el\hhsig}(\qr).
 \end{align}
Comparing the equations \re{csostrain} and \re{uregstrain} shows that
our method of time dependent elastic coefficients will lead to the same
temporal ordinary differential equations.
 
When we model the gradual pressurizing of the tube/tank, we multiply
the inner boundary condition by \m { \qlam(t) }. The corresponding
elastic solution is
 \begin{align}
\elt\hhsig^{} \71{t, \qr} & = \dev{\elt\hhsig} \71{t, \qr} +
\sph{\elt\hhsig} \71{t, \qr} =
 \qlam(t)\left[\dev{\el\hhsig}(\qr)+\sph{\el\hhsig}(\qr)\right],
 \\
\elt\hheps^{} \71{t, \qr} & = \f{1}{\dev{\qE}} \dev{\elt\hhsig} \71{t, \qr}
+ \f{1}{\sph{\qE}} \sph{\elt\hhsig} \71{t, \qr} =
\qlam(t) \left[ \f{1}{\dev{\qE}} \dev{\el\hhsig}(\qr) +
\f{1}{\sph{\qE}} \sph{\el\hhsig}(\qr) \right].
 \end{align}
For the rheological solution, the elasticity coefficients \m{\dev{\qE}}
and \m{\sph{\qE}} are changed to time dependent functions:
 \begingroup  \setlength{\jot}{8pt}
 \begin{align}
\rhe\hhsig \71{t, \qr} & = \qlam(t) \left[ \dev{\el\hhsig}(\qr) +
\sph{\el\hhsig}(\qr) \right],
 \\
\rhe\hheps \71{t, \qr} & = \qlam(t) \left[ \f{1}{\dev{\qE}(t)}
\dev{\el\hhsig}(\qr) + \f{1}{\sph{\qE}(t)} \sph{\el\hhsig}(\qr) \right] .
 \end{align}
 \endgroup
From this ansatz, the rheological operators generate the equations
 \begin{align}
 \dev{\oS}\qlam(t)=\dev{\oE}\f{\qlam(t)}{\dev{\qE}(t)},
 \qquad\qquad\qquad
 \sph{\oS}\qlam(t)=\sph{\oE}\f{\qlam(t)}{\sph{\qE}(t)},
 \end{align}
which can be solved for the two unknown functions \m { \dev{\qE}(t) },
\m { \sph{\qE}(t) } (or, more conveniently, for \m {
\f{\qlam(t)}{\dev{\qE}(t)} } and \m { \f{\qlam(t)}{\sph{\qE}(t)} }).
Again, initial conditions are taken from that,
for \mm{t<t_1}, \mm{\f{\qlam(t)}{\dev{\qE}(t)}=0} and
\mm{\f{\qlam(t)}{\sph{\qE}(t)}=0}.

\subsubsection{Cylindrical bore (tunnel) opened in an infinite homogeneous
but anisotropic stress field}  \label{anizot}

Now let us consider an infinite and homogeneous, but now anisotropic,
stress field, and let us analyize the rheological process caused by
drilling a cylindrical bore. This problem is a generalization of our
first example (cylindrical bore/tunnel opened in infinite, homogeneous
and isotropic stress field, Subsection~\ref{izotrop} with
\rf{outlineizo}). The solution of the elastic problem for fully opened
bore can be taken from \cite{anizoalagut},\footnote{Or can be checked
explicitly.} and is the sum of two terms containing linearly independent
space dependent functions,
---so to say, `spatial patterns'---,
the first term depending on \m{\qeta} and the
other being independent of it:
 \begingroup  \setlength{\jot}{8pt}
 \begin{align}
\el\hhsig(\qr)&=c(\eta)\hhsig_1(\qr)+\hhsig_2(\qr)
  \label{anizorug}
 = \f{1-\eta}{2+\eta}
 \begin{pmatrix}
0&0&0\\0&0&0\\0&0&-4\left(\f{R}{r}\right)^2\bsig_-(\qphi)
 \end{pmatrix}+
 \\  \tag*~
& +
 \begin{pmatrix}
\vspace{1.5mm}-\91{\f{R}{r}}^2\bsig_+ - \92{4\91{\f{R}{r}}^2-3\91{\f{R}{r}}^4}\bsig_-
(\qphi)&\92{2\91{\f{R}{r}}^2-3\91{\f{R}{r}}^4}\bsig_{r\qphi}(\qphi)&-\91{\f{R}{r}}^2\bsig_{rz}(\qphi)
 \\
  \vspace{1.5mm}
\92{2\91{\f{R}{r}}^2-3\91{\f{R}{r}}^4}\bsig_{r\qphi}(\qphi)&
\91{\f{R}{r}}^2\bsig_+
 -3\91{\f{R}{r}}^4\bsig_-(\qphi)&\91{\f{R}{r}}^2\bsig_{\qphi z}(\qphi)\\
-\91{\f{R}{r}}^2\bsig_{rz}(\qphi)&\91{\f{R}{r}}^2\bsig_{\qphi z}(\qphi)&0
 \end{pmatrix}
,
 \end{align}
 \endgroup
where we are using the following notations related to the primary field
\m{\bbsig}:
 \begin{align}
\bsig_+ & = \f{1}{2} \left(\bsig_{xx}+\bsig_{yy}\right),
 &
\bsig_-(\qphi) & = \f{1}{2} \left(\bsig_{xx}-\bsig_{yy}\right)
\cos{(2\qphi)}+\bsig_{xy}\sin{(2\qphi)},
 \nonumber  \\  \label{primfield}
\bsig_{rz}(\qphi) & =\bsig_{xz}\cos{\qphi}+\bsig_{yz}\sin{\qphi},
 &
\bsig_{r\qphi}(\qphi) & = -\f{1}{2}\left(\bsig_{xx}-\bsig_{yy}\right)
\sin{(2\qphi)}+\bsig_{xy}\cos{(2\qphi)},
 \\  \nonumber
\bsig_{\qphi z}(\qphi) & = -\bsig_{xz}\sin{\qphi}+\bsig_{yz}\cos{\qphi}.
& &
 \end{align}
Noticing that \m{\sph{\hhsig_1}(\qr)=\sph{\hhsig_2}(\qr)}, the strain
tensor can be written as
 \begingroup  \setlength{\jot}{8pt}
 \begin{align}
 \begin{split}
\el\hheps(\qr)&=\f{1}{\sph{\qE}}\left[\f{c(\eta)}{\eta}\dev{\hhsig_1}(\qr)+\f{1}{\eta}\dev{\hhsig_2}(\qr)+c(\eta)\sph{\hhsig_1}(\qr)+\sph{\hhsig_2}(\qr)\right]
\\
&=\f{1}{\sph{\qE}}\left\{\f{c(\eta)}{\eta}\dev{\hhsig_1}(\qr)+\f{1}{\eta}\dev{\hhsig_2}(\qr)+\left[c(\eta)+1\right]\sph{\hhsig_1}(\qr)\right\}
 \end{split}
 \end{align}
 \endgroup
[cf.\ \re{eta}].
Drilling is modelled again via multiplying the final boundary condition
by a
factor \m{\qlam(t)}; then the solution of
the elastic problem gets multiplied by the same \m{\qlam(t)}:
 \begingroup  \setlength{\jot}{8pt}
 \begin{align}
\elt\hhsig \71{t, \qr} & = \qlam(t) \23{
c(\eta)\hhsig_1(\qr)+\hhsig_2(\qr) },
 \\  \label{epsel}
\elt\hheps \71{t, \qr} & = \f{\qlam(t)}{\sph{\qE}}\left\{\f{c(\eta)}{\eta}\dev{\hhsig_1}(\qr)+\f{1}{\eta}\dev{\hhsig_2}(\qr)+\left[c(\eta)+1\right]\sph{\hhsig_1}(\qr)\right\}.
 \end{align}
 \endgroup
 
Following the already known recipe, let us replace the elasticity
coefficients in the stress and strain solution with unknown time
dependent functions. At this point we can realize that it is possible
(and will indeed be necessary) to substitute two separate function pairs
\m{\devs{\qE}\01{t}}, \m{\sphs{\qE}\01{t}} and \m{\deve{\qE}\01{t}},
\m{\sphe{\qE}\01{t}} [recall the argument in Section~\ref{method1}].
Accordingly, let us replace \m{\qeta} in the stress solution with
 \begin{align}
\eta_\qsig(t):=\f{\qEds(t)}{\qEss(t)},
 \end{align}
while in the strain solution we use some separate
time dependent function
 \begin{align}
\eta_\qeps(t):=\f{\qEde(t)}{\qEse(t)} ,
 \end{align}
in addition to
changing in \re{epsel} the explicite \m{\sph{\qE}} to \m{\qEse(t)}.

Since stress depends only on the ratio of \m{\qEds(t)} and \m{\qEss(t)}
[on \m { \eta_\qsig(t) } solely], there is only a one-function freedom
in stress. To fix the arbitrariness, we take the simplest choice
\mm{\qEss(t):= \qEs_0 } [which is a positive constant in case of solids;
cf.~\re{op2}].
 
We then have three functions to be determined. The rheological operators
generate three conditions on them, corresponding to the fact that the
strain solution \re{epsel} contains three linearly independent tensor
fields: \m{\dev{\hhsig_1}(\qr)}, \m{\dev{\hhsig_2}(\qr)} and
\m{\sph{\hhsig_1}(\qr)}. For the time dependent coefficient of each of
these independent tensor fields,
 \OMIT{%
(of these independent spatial
patterns\footnote{At this point, we can specify how to identify the
independent spatial patterns: they must have linearly independent \m {
\eta } dependent multipliers [cf. the curly brackets in \re{epsel}].}),
 }
one equation is generated; two equations follow from the deviatoric
rheological equation and one from the spherical one [cf.~\re{reoll}].
These three equations read
 \begin{align}  \label{nonlin_sode}
 \begin{split}
 \dev{\oS}\left[\qlam c(\eta_\qsig)\right]&=\dev{\oE}\left(\f{\qlam}{\qEse}\f{c(\eta_\qeps)}{\eta_\qeps}\right),\\
 \dev{\oS}\qlam&=\dev{\oE}\left(\f{\qlam}{\qEse}\f{1}{\eta_\qeps}\right),\\
 \sph{\oS}\left\{\qlam \left[c(\eta_\qsig)+1\right]\right\}&=\sph{\oE}\left\{\f{\qlam}{\qEse}\left[c(\eta_\qeps)+1\right]\right\}.
 \end{split}
 \end{align}
At first sight, solving this system of equations for \m { \eta_\qsig },
\m { \eta_\qeps }, \m { \qEse } seems difficult. Nevertheless,
introducing the auxiliary functions
 \begin{align}  \label{aux}
 \qlam_1:=\qlam c(\qetas),
 \qquad\qquad
 \kappa:=\f{\qlam}{\qetae\qEse},
 \qquad\qquad
 \kappa_1:=\f{c(\qetae)\qlam}{\qetae\qEse}
 \end{align}
and realizing the relationship
 \begin{align}
 c(\eta)+1=\f{1}{\eta} \22{1-2c(\eta)},
 \end{align}
one arrives at a system of \emph{linear} differential equations,
 \begin{align}
 \label{systode}
 \begin{split}
 \dev{\oS}\qlam_1 &= \dev{\oE}\kappa_1,\\
 \dev{\oS}\qlam &= \dev{\oE}\kappa,\\
 \sph{\oS}\left(\qlam_1+\qlam\right) &= \sph{\oE}\left(\kappa-2\kappa_1\right).
 \end{split}
 \end{align}
This system is solvable, and has a unique solution with, for example,
the assumption of `zero past' (zero complementary fields before the
opening, for a whole time interval).\footnote{Naturally, the original
functions \m{\qEds}, \m{\qEde} and \m{\qEse} are to be recovered from
\re{aux}.}
 
At this point, we can see that this method---the method of elasticity
constants made time dependent---is limited to three unknown time
dependent functions.

\subsection{Second approach: Method of elastic spatial patterns}
\label{second}

Here, we establish a method that
enables more than three unknown time dependent functions to be determined.
We assume that the solution of the elastic problem is of the form 
 \begin{align}  \label{53}
\el\hhsig(\qr)=\sum_{j=1}^J c_j(\qeta)\bitt\qs_j(\qr),
 \end{align}
where \m{J} is some integer, the coefficient functions
\m{c_j(\qeta)} are linearly independent of each other, and the spatial
patterns \m{\qs_j(\qr)} are also linearly independent. Notably,
 one could allow
the sum to be infinite; however, expansion with respect to an infinite
function series raises convergence questions, and here we wish to avoid
such mathematical complications. Fortunately, the finite sum form
already allows us to treat a good number of special cases, as shown
below.

In deviatoric--spherical separation, \re{53} reads
 \begin{align}
\el\hhsig(\qr)=\sum_{j=1}^J \left[c_j(\qeta)\dev{\qs}_j(\qr)+c_j(\qeta)\sph{\qs}_j(\qr)\right].
 \end{align}
Dimensional reasoning suggests to use, instead of strain,
a multiple of it that has the dimension of stress.
This can be simply achieved by\footnote{For solids, \m { \qEs }
is always positive.}
 \begin{align}  \label{feszdim_a}
\qqzet := \qEs \qqeps ,
 \end{align}
called hereafter stress-dimensioned strain.
Correspondingly,
 \begin{align}  \label{feszdim_al}
\hhzet = \qEs \hheps .
 \end{align}
The compatibility equation
\re{compp} remains in the same form for \m { \hhzet }:
 \begin{align}  \label{compzeta}
\nablar\times\hhzet\times\nablal=\zero .
 \end{align}
With \m { \hhzet }, Hooke's law is simplified to
 \begin{align}  \label{delgamzet}
\hhsig^{\rm d}
=\qeta\hhzet^{\rm d},
 \qquad
 \quad
\hhsig^{\rm s}
=\hhzet^{\rm s} .
 \end{align}
Then it is apparent that,
in a solution
of an elastic problem with stress boundary condition, \m { \hhzet }
depends on the elasticity coefficients only through \m{\qeta}, too [see
\re{eta}].
 
For the elastic solution, Hooke's law \re{hooke} leads to
 \begin{align}
\el\hhzet(\qr) = \sum_{j=1}^J \left[ \f{1}{\qeta}c_j
\8 1 {\qeta} \dev{\qs}_j(\qr) + c_j
\8 1 {\qeta} \sph{\qs}_j(\qr) \right] .
 \end{align}
When the stress boundary condition is multiplied by
a time dependent factor \m{\qlam\01{t}}, the corresponding
elastic stress and stress-dimensioned-strain solutions are multiplied
accordingly:
 \begin{align}
\elt\hhsig\71{t, \qr} & = \qlam(t)\sum_{j=1}^J \left[c_j(\qeta)\dev{\qs}_j(\qr)+c_j(\qeta)\sph{\qs}_j(\qr)\right],\\
\elt\hhzet\71{t, \qr} & = \qlam(t)\sum_{j=1}^J \left[
\f{1}{\qeta} c_j (\qeta) \dev{\qs}_j(\qr)+ c_j (\qeta) \sph{\qs}_j(\qr)\right].
 \end{align}
 
The rheological problem characterized by \re{reoll} imposes
 \begin{align}
\dev{\oS}\dev{\rhe\hhsig} \71{t, \qr} =
\dev{\oZ}\dev{\rhe\hhzet} \71{t, \qr},
 \qquad\qquad
\sph{\oS}\sph{\rhe\hhsig} \71{t, \qr} =
\sph{\oZ}\sph{\rhe\hhzet} \71{t, \qr}
 \end{align}
between stress and stress-dimensioned strain, where \12{\cf
\re{op1}--\re{op2}}
 \begin{align}  \label{71}
\dev{\oZ} = \qeta + \f{\qEd_1}{\qEs_0} \f{\pd}{\pd t} +
\f{\qEd_2}{\qEs_0} \f{\pd^2}{\pd t^2} + \cdots,
 \qquad
\sph{\oZ} = 1 + \f{\qEs_1}{\qEs_0} \f{\pd}{\pd t} +
\f{\qEs_2}{\qEs_0} \f{\pd^2}{\pd t^2} + \cdots
 \end{align}
(recall that, as seen in Sections~\ref{elrheol} and \ref{anmeth}, \m
{\qEs_0 } plays the role of \m { \qEs } in rheology).
 
As in the previous method, we look for the rheological solution as a
time parametrized succession of elastic solutions, in order to satisfy
the spatial conditions \re{impp}, \re{compzeta} and the stress boundary
condition. However, while previously \m { \eta } was replaced with time
dependent functions, now let us fix \m{J} constants \m{ \qeta_1,
\qeta_2, \ldots, \qeta_k, \ldots, \qeta_J}, consider the corresponding
elastic solutions, and take a time dependent linear combination of them
as our new ansatz. We allow for separate time dependent coefficients for
\m { \hhsig } and \m { \hhzet }:
 \begin{align}  \label{reolsig}
\rhe{\hhsig} \71{t, \qr} & = \sum_{k=1}^J \left\{ \qlam_k(t) \sum_{j=1}^J
\left[c_j\81{\qeta_k} \dev{\qs}_j (\qr) + c_j \81{\qeta_k}
\sph{\qs}_j(\qr)\right]\right\},
 \\  \label{reolzeta}
\rhe{\hhzet} \71{t, \qr} & = \sum_{k=1}^J \left\{ \qkap_k(t)
\sum_{j=1}^J \left[ \f{1}{\qeta_k} c_j \81{\qeta_k} \dev{\qs}_j(\qr)
+c_j \81{ \qeta_k } \sph{\qs}_j(\qr)\right]\right\}.
 \end{align}
The explanation for why to use exactly \m{J}
elastic solutions is that we have \m{J}
linearly independent
spatial patterns, the vector space of spatial patterns is \m{J}
dimensional, so using more than \m{J} pieces of combinations would be
redundant, would not increase our freedom.

Thanks to working from elastic solutions, the spatial equations
\re{impp} and \re{compzeta} are satisfied, while satisfying the
boundary condition is ensured via the condition
 \begin{align}  \label{sumlam}
\sum_{k=1}^J\qlam_k(t)=\qlam(t) .
 \end{align}
 
What is left is only to fulfil the rheological relationships \re{71}.
From \re{reolsig}--\re{reolzeta}, one finds
 \begin{align}  \label{63}
\dev{\oS}\left\{\sum_{k=1}^J\left[\qlam_k(t)\sum_{j=1}^J c_j
\81{\qeta_k} \dev{\qs}_j(\qr)\right]\right\} & =
\dev{\oZ}\left\{\sum_{k=1}^J\left[\qkap_k(t)\sum_{j=1}^J \f{1}{\qeta_k}
c_j \81{\qeta_k} \dev{\qs}_j(\qr)\right]\right\},
 \\  \label{64}
\sph{\oS}\left\{\sum_{k=1}^J\left[\qlam_k(t)\sum_{j=1}^J c_j
\81{\qeta_k} \sph{\qs}_j(\qr)\right]\right\} & =
\sph{\oZ}\left\{\sum_{k=1}^J\left[\qkap_k(t)\sum_{j=1}^J c_j
\81{\qeta_k} \sph{\qs}_j(\qr)\right]\right\}.
 \end{align}
An important advantage of this second method is that, apparently, the
system of differential equations to be solved for the functions
\m{\qlam_k(t)}, \m{\qkap_k(t)} is \emph{linear}.
 

The question arises whether the solution depends on the choice of the
values \m{\qeta_1,\ldots,\qeta_J}. Now, the \m{J}
elastic solutions form
a basis\footnote{Apart probably from certain special choices of
\m{\qeta_1,\ldots,\qeta_J}.} in the space of the linear combinations of
the \m{J} spatial patterns. Therefore, another set of values
\m{\qeta_1,\ldots,\qeta_J} represents another basis, corresponding to
which the linear expansion coefficients of the solution are some other
functions \m{\qlam_k(t)}, \m{\qkap_k(t)}. The solution itself is the
same (assuming the appropriate amount of initial conditions, naturally).
 
To simplify the notations, let us introduce the matrix
 \begin{align}  \label{65}
 C_{jk}=c_j(\qeta_k)
 \end{align}
of constant elements. This matrix is nondegenerate---except probably for
certain special choices of \m{\qeta_1,\ldots,\qeta_J}---, since
the functions \m{c_j(\qeta)} are also linearly independent.
 
Formulae \re{63}--\re{64} can be rewritten as
 \begin{align}
 \label{66}
 \sum_{j=1}^J\left[\sum_{k=1}^J\dev{\oS}\qlam_k(t) C_{jk}\dev{\qs}_j(\qr)\right]&=
 \sum_{j=1}^J\left[\sum_{k=1}^J\dev{\oZ}\qkap_k(t) \f{1}{\qeta_k}C_{jk}\dev{\qs}_j(\qr)\right],\\
 \label{67}
 \sum_{j=1}^J\left[\sum_{k=1}^J\sph{\oS}\qlam_k(t) C_{jk}\sph{\qs}_j(\qr)\right]&=
 \sum_{j=1}^J\left[\sum_{k=1}^J\sph{\oZ}\qkap_k(t) C_{jk}\sph{\qs}_j(\qr)\right].
 \end{align}
The spatial patterns \m{\qs_1(\qr),\ldots,\qs_J(\qr)} are linearly
independent functions. If their deviatoric parts are also linearly
independent and their spherical parts are also linearly independent,
\re{66}--\re{67} require equality of the coefficients of each component:
 \begin{align}  \label{68}
 &&
\sum_{k=1}^J\dev{\oS}\qlam_k(t) C_{jk}&=
\sum_{k=1}^J\dev{\oZ}\qkap_k(t) \f{1}{\qeta_k}C_{jk},
 &j&=1,\ldots,J,
 &&
 \\  \label{69}
 &&
\sum_{k=1}^J\sph{\oS}\qlam_k(t) C_{jk}&=
\sum_{k=1}^J\sph{\oZ}\qkap_k(t) C_{jk},
 &j&=1,\ldots,J.
 &&
 \end{align}
These are \m{2J} equations, and taking \re{sumlam} also into account,
altogether we have \m{2J+1} equations for the \m{2J} unknown functions
\m{ \qlam_1(t), \ldots, \qlam_J(t), \qkap_1(t), \ldots, \qkap_J(t)}.
Hence, in general, this second method cannot provide a solution.
Hovewer, we can observe in each of the problems analyzed subsequently
that the \m{J} spherical patterns
\m{\sph{\qs}_1(\qr),\ldots,\sph{\qs}_J(\qr)} are not linearly
independent but one of them can be expressed as a linear comination of
the others. Then \re{67} means only \m{J-1} independent equations and
the method can lead to a solution. If the initial conditions can also be
written in the form of \re{reolsig}--\re{reolzeta} -- \eg the `zero past
history' assumption%
is of this form, corresponding to \m{\qlam = 0} --, then the found
solution is \emph{the} solution of the problem.
 
Let us now see some examples how this method works and performs in
practice. The first three problems
have already been solved with the method of time dependent elastic
coefficients, too, so comparison can also be made.

\subsubsection{Cylindrical bore (tunnel) and spherical hollow opened in
infinite, homogeneous and isotropic stress field}  \label{izotropp}

As the first example, let us consider a cylindrical bore and a spherical
hollow opened in an infinite, homogeneous and isotropic stress field. As
happened in Subsections~\ref{izotrop}--\ref{hollow}, these two problems
are expected to lead to the same rheological ordinary differential
equations to be solved.

The elastic stress solutions are in this case traceless---pure
deviatoric---tensors [see Subsections~\ref{izotrop}--\ref{hollow}],
which can be written, in the notation of \re{53}, as 
 \begin{align}
\el\hhsig(\qr)=c \81{\qeta} \qs(\qr),
 \end{align}
where \m{c(\qeta)=1} and \m{\qs(\qr)=\hhsig(\qr)}, while the
stress-dimensioned strain is
 \begin{align}
\el\hhzet(\qr) = \f{1}{\qEd} \dev{\el\hhsig}(\qr) =
\f{1}{\qEd} \dev{\qs}(\qr) = \dev{\el\hhzet}(\qr).
 \end{align}
The elastic solution of both problems, with time dependent boundary
conditions, is
 \begin{align}
\elt\hhsig \71{t, \qr} & = \qlam(t)\cdot\dev{\qs}(\qr),
 \\
\elt\hhzet \71{t, \qr} & = \f{\qlam(t)}{\qeta}\cdot\dev{\qs}(\qr).
 \end{align} 
According to the second method, the rheological solution is looked for
in the form of
 \begin{align}
\rhe{\hhsig} \71{t, \qr} & = \qlam(t)\cdot\dev{\qs}(\qr),\\
\rhe{\hhzet} \71{t, \qr} & = \f{\qkap(t)}{\qeta}\cdot\dev{\qs}(\qr)
 \end{align} 
[cf.~\re{reolsig}--\re{reolzeta}]. The spherical part related equation
\re{69} is satisfied trivially, while the deviatoric condition \re{69}
generates
 \begin{align}
 \label{80}
 \dev{\oS}\qlam(t)=\dev{\oZ}\f{\qkap(t)}{\qeta}.
 \end{align}
Comparing this with \re{izoalagutreol} and \re{gomb} shows that we have
reached the same equations.

\subsubsection{Cylindrical bore (tunnel) opened in infinite, homogeneous
but anisotropic stress field}

The elastic solution of this problem, \re{anizorug}, can be written in
the form of \re{53} as
 \begingroup  \setlength{\jot}{8pt}
 \begin{align}
\el\hhsig(\qr) & = c_1 \81{\eta} \qs_1(\qr)+c_2 \81{\qeta} \qs_2(\qr)=
 \nonumber  \\
&=
 \begin{pmatrix}
\vspace{1.5mm}
-\91{\f{R}{r}}^2\bsig_+ -
\92{4\91{\f{R}{r}}^2-3\91{\f{R}{r}}^4}\bsig_-(\qphi)&
\92{2\91{\f{R}{r}}^2-3\91{\f{R}{r}}^4}\bsig_{r\qphi}
(\qphi)&-\91{\f{R}{r}}^2\bsig_{rz}(\qphi)
 \\
 \vspace{1.5mm}
\92{2\91{\f{R}{r}}^2-3\91{\f{R}{r}}^4}\bsig_{r\qphi}(\qphi) &
\91{\f{R}{r}}^2\bsig_+ -3\91{\f{R}{r}}^4\bsig_-(\qphi) &
\91{\f{R}{r}}^2\bsig_{\qphi z}(\qphi)
 \\
-\91{\f{R}{r}}^2\bsig_{rz}(\qphi) &
\91{\f{R}{r}}^2\bsig_{\qphi z}(\qphi)&0
 \end{pmatrix} +
 \nonumber  \\  \label{anizosig}
&
\quad
 +\f{1-\eta}{2+\eta}
 \begin{pmatrix}
0&0&0\\0&0&0\\0&0&-4\left(\f{R}{r}\right)^2\bsig_-(\qphi)
 \end{pmatrix} ,
 \end{align}
 \endgroup
where \m{c_1(\qeta)=1}, and the notations for \m{\bbsig} can be found in
\re{primfield}.
 
Since we have two independent spatial patterns, we are looking for two
functions \m{\qkap_k(t)} and two functions \m{\qlam_k(t)}, and
the introduced matrix \m{C_{jk}} in \re{65} reads here
 \begin{align}
 C_{jk}=\begin{pmatrix}1&1\\ \displaystyle{\f{1-\qeta_1}{2+\qeta_1}}&\displaystyle{\f{1-\qeta_2}{2+\qeta_2}}\end{pmatrix}.
 \end{align}
Consequently, the rheological solution is looked for, based on
\re{reolsig}--\re{reolzeta}, in the form
 \begin{align}  \label{sigrug}
 \begin{split}
\rhe{\hhsig} \71{t, \qr} & = \qlam_1(t)C_{11}\dev{\qs}_1(\qr)+\qlam_1(t)C_{21}\dev{\qs}_2(\qr)+\qlam_2(t)C_{12}\dev{\qs}_1(\qr)+\qlam_2(t)C_{22}\dev{\qs}_2(\qr)\mathrel{+}
 \\
&
 \quad
+\qlam_1(t)C_{11}\sph{\qs}_1(\qr)+\qlam_1(t)C_{21}\sph{\qs}_2(\qr)+\qlam_2(t)C_{12}\sph{\qs}_1(\qr)+\qlam_2(t)C_{22}\sph{\qs}_2(\qr)
 \end{split}\\  \label{zetrug}
 \begin{split}
\rhe{\hhzet} \71{t, \qr} & = \qkap_1(t)\f{C_{11}}{\qeta_1}\dev{\qs}_1(\qr)+\qkap_1(t)\f{C_{21}}{\qeta_1}\dev{\qs}_2(\qr)+\qkap_2(t)\f{C_{12}}{\qeta_2}\dev{\qs}_1(\qr)+\qkap_2(t)\f{C_{22}}{\qeta_2}\dev{\qs}_2(\qr)\mathrel{+}
 \\
&
 \quad
+\qkap_1(t)C_{11}\sph{\qs}_1(\qr)+\qkap_1(t)C_{21}\sph{\qs}_2(\qr)+\qkap_2(t)C_{12}\sph{\qs}_1(\qr)+\qkap_2(t)C_{22}\sph{\qs}_2(\qr).
 \end{split}
 \end{align}
As we have noticed in Subsection~\ref{anizot},
\m{\sph{\qs}_1=\sph{\qs}_2}. Applying this and imposing the rheological
requirements on \re{sigrug} and \re{zetrug} leads to the equations
 \begin{align}
 \nonumber
 \dev{\oS}\left[\qlam_1(t)C_{11}+\qlam_2(t)C_{12}\right]&=\dev{\oZ}\left[\qkap_1(t)\f{C_{11}}{\qeta_1}+\qkap_2(t)\f{C_{12}}{\qeta_2}\right],\\
 \dev{\oS}\left[\qlam_1(t)C_{21}+\qlam_2(t)C_{22}\right]&=\dev{\oZ}\left[\qkap_1(t)\f{C_{21}}{\qeta_1}+\qkap_2(t)\f{C_{22}}{\qeta_2}\right],\\
 \nonumber
 \sph{\oS}\left\{\qlam_1(t)\left[C_{11}+C_{21}\right]+\qlam_2(t)\left[C_{12}+C_{22}\right]\right\}&=
 \sph{\oZ}\left\{\qkap_1(t)\left[C_{11}+C_{21}\right]+\qkap_2(t)\left[C_{12}+C_{22}\right]\right\} .
 \end{align}
For the four unknowns, besides these three equations, the
boundary condition related \re{sumlam} gives the fourth equation:
 \begin{align}
 \qlam(t)=\qlam_1(t)+\qlam_2(t).
 \end{align}

Solutions for concrete rheological models can be found in
Section~\ref{concrete}.

\subsubsection{Cylindrical bore (tunnel) opened in homogeneous medium
loaded by its self weight}

As our most complicated example, now we discuss the rheological process
of a semi-infinite domain loaded by its own weight and weakened by a
bore. To describe the rheological process of an underground tunnel, this
model can be considered as more accurate than the previous bore models.

 \begin{figure}[H]  \centering
   \ig[width=0.4\textwidth]{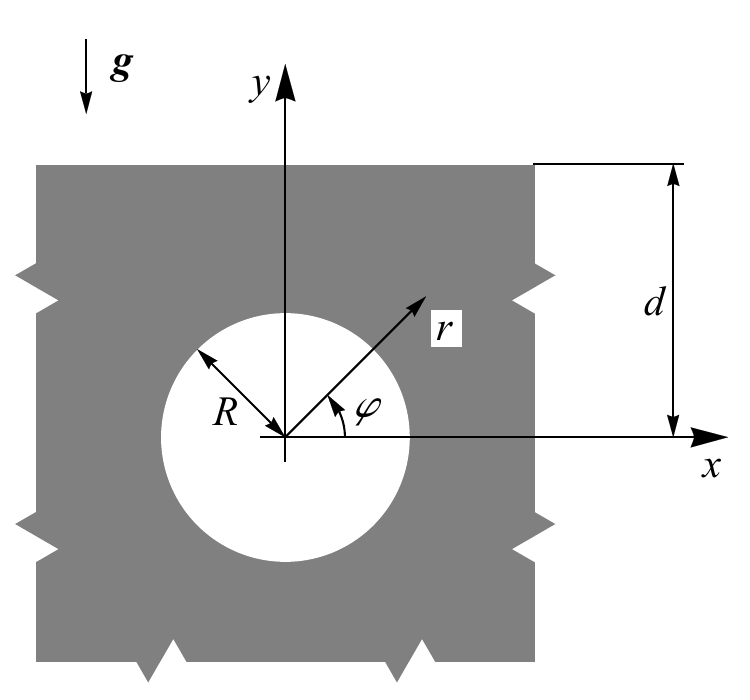}
\caption{Outline and notations for the cylindrical bore opened in
homogeneous medium loaded by its self weight.}
  \label{outlineself}
 \end{figure}

The elastic solution of the problem can be found in \cite{mindlin,
papkovich, savin}. The primary stress field---the one before the
drilling---, in
an appropriate Cartesian coordinate system
(see \rf{outlineself}), can be written as
 \begin{align}  \label{primer}
\bbsig=\gamma(y-d)\begin{pmatrix}k&0&0\\0&1&0\\0&0&k\end{pmatrix},
 \end{align}
where \m{\gamma=\qrho g} describes the homogeneous force density, \m{d}
is the depth of the center of the bore from the surface and the parameter
\m{k} is called lateral pressure factor. \cite{mindlin} gives the
solution of the problem for three different values of \m{k}:
 \begin{itemize}
 \item
When we assume hydrostatic pressure distribution for the primary field
then \m{k=1}. This is a good approximation for tunnels opened at large
depths.
 \item
When one can assume that the dilatation of the medium is laterally
inhibited then the strain components \m{\beps_{xx}} and \m{\beps_{zz}}
are zeros so one can derive for the lateral pressure factor
\m{k=\f{\nu}{1-\nu}=\f{1-\qeta}{1+2\qeta}}.
 \item
When one can assume that the dilatation of the medium is laterally
free then the stress components \m{\bsig_{xx}} and \m{\bsig_{zz}} are
zeros so \m{k=0}.
 \end{itemize}

Transforming \re{primer} to cylindrical coordinate system yields the
stress components
 \begin{align}
 \begin{split}
 \bsig_{rr}&=\f{\gamma r}{4}\left[\left(3+k\right)\sin{\qphi}+\left(k-1\right)\sin{3\qphi}\right]-\f{\gamma d}{2}\left[\left(1+k\right)-\left(1-k\right)\cos{2\qphi}\right],\\
 \bsig_{\qphi\qphi}&=\f{\gamma r}{4}\left[\left(1+3k\right)\sin{\qphi}-\left(k-1\right)\sin{3\qphi}\right]-\f{\gamma d}{2}\left[\left(1+k\right)+\left(1-k\right)\cos{2\qphi}\right],\\
 \bsig_{r\qphi}&=\f{\gamma r}{4}\left(1-k\right)\left(\cos{\qphi}-\cos{3\qphi}\right)-\f{\gamma d}{2}\left(1-k\right)\sin{2\qphi},\\
 \bsig_{zz}&=\gamma r k \sin{\qphi}-\gamma d k,\\
 \bsig_{rz}&=\bsig_{\qphi z}=0 .
 \end{split}
 \end{align}
 
The boundary conditions are prescribed for the contour of the cylinder
and for the plane surface---the horizontal boundary---after the
drilling; on these boundaries the normal component of stress is
zero.

Mindlin gives the solution in form of an infinite series in bipolar
coordinate system -- which suits to both the cylinder and the plane
\cite{mindlin}. If
the ratio of the depth \m{d} of the
center of the bore from the surface and the radius \m{R} of the cylinder
satisfies \m{\f{d}{R}>1.5}---large-depth approximation---, then
it suffices to take the leading order term from the infinite series.

This first term is transformed to cylindrical coordinate system
in \cite {papkovich, savin}:
 \begin{align}
\qsig_{rr}&=\f{\gamma R}{4}\left\{\left[\left(3+k\right)\f{r}{R}-\f{4+5\qeta}{1+2\qeta}\f{R}{r}+\left(\f{1-\qeta}{1+2\qeta}-k\right)\f{R^3}{r^3}\right]\sin{\qphi}\right.+\nonumber\\&\qquad+\left.\left[\left(k-1\right)\f{r}{R}+5\left(1-k\right)\f{R^3}{r^3}+4\left(k-1\right)\f{R^5}{r^5}\right]\sin{3\qphi}\right\}-\nonumber\\&\qquad-\f{\gamma d}{2}\left[\left(1+k\right)\left(1-\f{R^2}{r^2}\right)+\left(1-k\right)\left(-1+4\f{R^2}{r^2}-3\f{R^4}{r^4}\right)\cos{2\qphi}\right],
 \\
\qsig_{\qphi\qphi}&=\f{\gamma R}{4}\left\{\left[\left(1+3k\right)\f{r}{R}+\f{3\qeta}{1+2\qeta}\f{R}{r}+\left(k-\f{\qeta}{1+2\qeta}\right)\f{R^3}{r^3}\right]\sin{\qphi}\right.+\hskip 3.6em\nonumber\\&\qquad+\left.\left[\left(1-k\right)\f{r}{R}+\left(k-1\right)\f{R^3}{r^3}+4\left(1-k\right)\f{R^5}{r^5}\right]\sin{3\qphi}\right\}-\nonumber\\&\qquad-\f{\gamma d}{2}\left[\left(1+k\right)\left(1+\f{R^2}{r^2}\right)+\left(1-k\right)\left(1+3\f{R^4}{r^4}\right)\cos{2\qphi}\right],
 \\
\qsig_{r\qphi}&=\f{\gamma R}{4}\left\{\left[\left(1-k\right)\f{r}{R}-\f{3\qeta}{1+2\qeta}\f{R}{r}+\left(k-\f{\qeta}{1+2\qeta}\right)\f{R^3}{r^3}\right]\cos{\qphi}\right.+\hskip 1.1em\nonumber\\&\qquad+\left.\left[\left(k-1\right)\f{r}{R}+3\left(k-1\right)\f{R^3}{r^3}+4\left(1-k\right)\f{R^5}{r^5}\right]\cos{3\qphi}\right\}-\nonumber\\&\qquad-\f{\gamma d}{2}\left(1-k\right)\left(1+2\f{R^2}{r^2}-3\f{R^4}{r^4}\right)\sin{2\qphi},
 \\
\qsig_{zz}&=\f{\gamma R}{4}\left[\left(4k\f{r}{R}-2\f{1-\qeta}{1+2\qeta}\f{R}{r}\right)\sin{\qphi}+4\f{1-\qeta}{2+\qeta}\left(1-k\right)\f{R^3}{r^3}\sin{3\qphi}\right]-\nonumber\\&\qquad-\f{\gamma d}{2}\left(2k+4\f{1-\qeta}{2+\qeta}\f{R^2}{r^2}\cos{2\qphi}\right),
 \\
\qsig_{rz} &= \qsig_{\qphi z}=0
 \end{align}
The entries of the complementary field
(having a plane strain situation)
are
 \begin{align}
\hsig_{rr}&=\f{\gamma R}{4}\left\{\left[-\f{4+5\qeta}{1+2\qeta}\f{R}{r}+\left(\f{1-\qeta}{1+2\qeta}-k\right)\f{R^3}{r^3}\right]\sin{\qphi}\right.+\hskip 4.4 em\nonumber\\&\qquad+\left.\left[5\left(1-k\right)\f{R^3}{r^3}+4\left(k-1\right)\f{R^5}{r^5}\right]\sin{3\qphi}\right\}-\nonumber\\&\qquad-\f{\gamma d}{2}\left[-\left(1+k\right)\f{R^2}{r^2}+\left(1-k\right)\left(4\f{R^2}{r^2}-3\f{R^4}{r^4}\right)\cos{2\qphi}\right],
 \\
\hsig_{\qphi\qphi}&=\f{\gamma R}{4}\left\{\left[\f{3\qeta}{1+2\qeta}\f{R}{r}+\left(k-\f{\qeta}{1+2\qeta}\right)\f{R^3}{r^3}\right]\sin{\qphi}\right.+\hskip 5.1 em\nonumber\\&\qquad+\left.\left[\left(k-1\right)\f{R^3}{r^3}+4\left(1-k\right)\f{R^5}{r^5}\right]\sin{3\qphi}\right\}-\nonumber\\&\qquad-\f{\gamma d}{2}\left[\left(1+k\right)\f{R^2}{r^2}+3\left(1-k\right)\f{R^4}{r^4}\cos{2\qphi}\right],
 \\
\hsig_{r\qphi}&=\f{\gamma R}{4}\left\{\left[-\f{3\qeta}{1+2\qeta}\f{R}{r}+\left(k-\f{\qeta}{1+2\qeta}\right)\f{R^3}{r^3}\right]\cos{\qphi}\right.+\hskip 4.2 em\nonumber\\&\qquad+\left.\left[3\left(k-1\right)\f{R^3}{r^3}+4\left(1-k\right)\f{R^5}{r^5}\right]\cos{3\qphi}\right\}-\nonumber\\&\qquad-\f{\gamma d}{2}\left(1-k\right)\left(2\f{R^2}{r^2}-3\f{R^4}{r^4}\right)\sin{2\qphi},
 \\
\hsig_{zz}&=\f{1-\qeta}{1+2\qeta}\cdot\left\{\f{\gamma R}{4}\left[-2\f{2+\qeta}{1+2\qeta}\f{R}{r}\sin{\qphi}+4\left(1-k\right)\f{R^3}{r^3}\sin{3\qphi}\right]\right.-\nonumber\\&\qquad-\left.2\gamma d\left(1-k\right)\f{R^2}{r^2}\cos{2\qphi}\right\},
 \\
\hsig_{rz}&= \hsig_{\qphi z}=0.
 \end{align}

Now we analyze the rheological process caused by the drilling for the
above-mentioned three values of the lateral pressure factor.

\subsubsection*{Hydrostatic initial stress state (\m{k=1})}

In this case, the complementary field can be written in the form of
\re{53} as
 \begin{align}  \label{sighidro}
 \begin{split}
\el\hhsig(\qr)&=c_1 \81{\qeta} \qs_1(\qr)+c_2 \81{\qeta} \qs_2(\qr)=
 \\[1ex]
& = \begin{pmatrix}
-\f{\gamma R}{4} \left( 3\f{R}{r}+\f{R^3}{r^3} \right) \sin{\qphi} +
\gamma d \f{R^2}{r^2} & \f{\gamma R}{4} \left( -\f{R}{r}+\f{R^3}{r^3}
\right) \cos{\qphi} & 0
 \\[1.3ex]
\f{\gamma R}{4} \left( -\f{R}{r} + \f{R^3}{r^3} \right) \cos{\qphi} &
\f{\gamma R}{4} \left( \f{R}{r} + \f{R^3}{r^3} \right) \sin{\qphi} -
\gamma d \f{R^2}{r^2} & 0
 \\[1.3ex]
0 & 0 & 0
 \end{pmatrix}
\mathrel{+}
 \\[1ex]
& \qquad+\f{1-\qeta}{1+2\qeta} \begin{pmatrix}
\f{\gamma R}{4}\left(-\f{R}{r}+\f{R^3}{r^3}\right)\sin{\qphi}&
\f{\gamma R}{4}\left(\f{R}{r}-\f{R^3}{r^3}\right)\cos{\qphi}& 0
 \\[1.3ex]
\f{\gamma R}{4}\left(\f{R}{r}+\f{R^3}{r^3}\right)\cos{\qphi} &
-\f{\gamma R}{4}\left(\f{R}{r}+\f{R^3}{r^3}\right)\sin{\qphi} & 0
 \\[1.3ex]
0&0&-\f{\gamma R}{2}\f{R}{r}\sin{\qphi}\end{pmatrix},
 \end{split}
 \end{align}
here \m{c_1(\qeta)=1} again so the matrix \m{C_{jk}} is
 \begin{align}
 C_{jk}=\begin{pmatrix}1&1\\\displaystyle{\f{1-\qeta_1}{1+2\qeta_1}}&\displaystyle{\f{1-\qeta_2}{1+2\qeta_2}}\end{pmatrix}.
 \end{align}
The rheological solution is looked for in the form of
\re{reolsig}--\re{reolzeta} as
 \begin{align}  \label{sigrug1}
 \begin{split}
\rhe{\hhsig} \71{t, \qr} & = \qlam_1(t)C_{11}\dev{\qs}_1(\qr)+\qlam_1(t)C_{21}\dev{\qs}_2(\qr)+\qlam_2(t)C_{12}\dev{\qs}_1(\qr)+\qlam_2(t)C_{22}\dev{\qs}_2(\qr)\mathrel{+}
 \\
&
 \quad
+\qlam_1(t)C_{11}\sph{\qs}_1(\qr)+\qlam_1(t)C_{21}\sph{\qs}_2(\qr)+\qlam_2(t)C_{12}\sph{\qs}_1(\qr)+\qlam_2(t)C_{22}\sph{\qs}_2(\qr),
 \end{split}\\  \label{zetrug1}
 \begin{split}
\rhe{\hhzet} \71{t, \qr} & = \qkap_1(t)\f{C_{11}}{\qeta_1}\dev{\qs}_1(\qr)+\qkap_1(t)\f{C_{21}}{\qeta_1}\dev{\qs}_2(\qr)+\qkap_2(t)\f{C_{12}}{\qeta_2}\dev{\qs}_1(\qr)+\qkap_2(t)\f{C_{22}}{\qeta_2}\dev{\qs}_2(\qr)\mathrel{+}
 \\
&
 \quad
+\qkap_1(t)C_{11}\sph{\qs}_1(\qr)+\qkap_1(t)C_{21}\sph{\qs}_2(\qr)+\qkap_2(t)C_{12}\sph{\qs}_1(\qr)+\qkap_2(t)C_{22}\sph{\qs}_2(\qr).
 \end{split}
 \end{align}
Noticing that \m{2\sph{\qs}_1=\sph{\qs}_2}, there will be one spherical
equation less than deviatoric:
 \begin{align}
 \nonumber\dev{\oS}\big[\qlam_1(t)C_{11}+\qlam_2(t)C_{12}\big]&=\dev{\oZ}\left[\qkap_1(t)\f{C_{11}}{\qeta_1}+\qkap_2(t)\f{C_{12}}{\qeta_2}\right],\\
 \dev{\oS}\big[\qlam_1(t)C_{21}+\qlam_2(t)C_{22}\big]&=\dev{\oZ}\left[\qkap_1(t)\f{C_{21}}{\qeta_1}+\qkap_2(t)\f{C_{22}}{\qeta_2}\right],\\
 \sph{\oS}\Big\{\biTT\qlam_1(t)\biTTTT\left[C_{11}+2C_{21}\right]+\qlam_2(t)\biTTTT\left[C_{12}+2C_{22}\right]\biTT\Big\}&=
 \nonumber\sph{\oZ}\Big\{\biTT\qkap_1(t)\biTTTT\left[C_{11}+2C_{21}\right]+\qkap_2(t)\biTTTT\left[C_{12}+2C_{22}\right]\biTT\Big\};
 \end{align}
 and, from the boundary condition,
 \begin{align}
 \qlam(t)=\qlam_1(t)+\qlam_2(t).
 \end{align}

\subsubsection*{No lateral deformations
(\m{k=\f{\nu}{1-\nu}=\f{1-\qeta}{1+2\qeta}})}

Now the complementary stress field is
 \begin{align}  \label{signolat}
\el\hhsig(\qr)&=c_1 \81{\qeta} \qs_1(\qr)+
c_2 \81{\qeta} \qs_2(\qr)+c_3 \81{\qeta} \qs_3(\qr),
 \end{align}
 where
 \begin{align}
&&
c_1(\qeta)&=\f{1-\qeta}{1+2\qeta},&
c_2(\qeta)&=\f{2+\qeta}{1+2\qeta},&
c_3(\qeta)&=\f{3(1-\qeta)\qeta}{(2+\qeta)(1+2\qeta)},
&&
 \end{align}
and
 \begin{align}
 \nonumber
 s_{1,rr}(\qr)&=\f{\gamma R}{2}\left[\f{R}{r}\sin{\qphi}-\left(5\f{R^3}{r^3}-4\f{R^5}{r^5}\right)\sin{3\qphi}\right]+\gamma d \left(4\f{R^2}{r^2}-3\f{R^4}{r^4}\right)\cos{2\qphi},\\
 \nonumber
 s_{1,\qphi\qphi}(\qr)&=\f{\gamma R}{2}\left[-\f{R}{r}\sin{\qphi}+\left(\f{R^3}{r^3}-4\f{R^5}{r^5}\right)\sin{3\qphi}\right]+3\gamma d \f{R^4}{r^4}\cos{2\qphi},\\
 \nonumber
 s_{1,r\qphi}(\qr)&=\f{\gamma R}{2}\left[\f{R}{r}\cos{\qphi}+\left(3\f{R^3}{r^3}-4\f{R^5}{r^5}\right)\cos{3\qphi}\right]+\gamma d \left(2\f{R^2}{r^2}-3\f{R^4}{r^4}\right)\sin{2\qphi},\\
 \nonumber
 s_{1,zz}(\qr)&=-\f{\gamma R}{2}\cdot\f{R}{r}\sin{\qphi},
 \qquad
 s_{1,rz}(\qr)=s_{1,\qphi z}(\qr)=0,\\
 \nonumber
 s_{2,rr}(\qr)&=\f{\gamma R}{4}\biTTTT\left[\biTT-3\f{R}{r}\sin{\qphi}+\biTTTT\left(5\f{R^3}{r^3}-4\f{R^5}{r^5}\right)\biTTTT\sin{3\qphi}\biTT\right]\biTTTT-\f{\gamma d}{2}\biTTTT\left[\left(4\f{R^2}{r^2}-3\f{R^4}{r^4}\right)\biTTTT\cos{2\qphi}-\f{R^2}{r^2}\right]\biTT,\\
 \nonumber
 s_{2,\qphi\qphi}(\qr)&=\f{\gamma R}{4}\left[\f{R}{r}\sin{\qphi}-\left(\f{R^3}{r^3}-4\f{R^5}{r^5}\right)\sin{3\qphi}\right]-\f{\gamma d}{2}\left(3\f{R^4}{r^4}\cos{2\qphi}+\f{R^2}{r^2}\right),\\
 \nonumber
 s_{2,r\qphi}(\qr)&=\f{\gamma R}{4}\left[-\f{R}{r}\cos{\qphi}-\left(3\f{R^3}{r^3}+4\f{R^5}{r^5}\right)\cos{3\qphi}\right]-\f{\gamma d}{2}\left(2\f{R^2}{r^2}-3\f{R^4}{r^4}\right)\sin{2\qphi},\\
 \nonumber
 s_{2,zz}(\qr)&=s_{2,rz}(\qr)=s_{2,\qphi z}(\qr)=0,\\
 \qs_3(\qr)&=\begin{pmatrix}0&0&0\\0&0&0\\0&0&\gamma R\f{R^3}{r^3}\sin{3\qphi}-2\gamma d
\f{R^2}{r^2}\cos{2\qphi}\end{pmatrix}.
 \end{align}
One finds that \mm{\sph{\qs}_2=\sph{\qs}_1+3\sph{\qs}_3} so this time
there is again one spherical equation less than deviatoric. In this
case, the matrix \m{C_{jk}} is
 \begin{align}  
C_{jk} = \begin{pmatrix}
\displaystyle{\f{1-\qeta_1}{1+2\qeta_1}} &
\displaystyle{\f{1-\qeta_2}{1+2\qeta_2}} &
\displaystyle{\f{1-\qeta_3}{1+2\qeta_3}}
 \\[3ex]
\displaystyle{\f{2+\qeta_1}{1+2\qeta_1}} &
\displaystyle{\f{2+\qeta_2}{1+2\qeta_2}} &
\displaystyle{\f{2+\qeta_3}{1+2\qeta_3}}
 \\[3ex]
\displaystyle{\f{3(1-\qeta_1)\qeta_1}{(2+\qeta_1)(1+2\qeta_1)}} &
\displaystyle{\f{3(1-\qeta_2)\qeta_2}{(2+\qeta_2)(1+2\qeta_2)}} &
\displaystyle{\f{3(1-\qeta_3)\qeta_3}{(2+\qeta_3)(1+2\qeta_3)}}
 \end{pmatrix}.
 \end{align}
 The rheological solution is looked for in the form of
 \begin{align}
 \label{sigrug2}
 \begin{split}
\rhe{\hhsig} \71{t, \qr} & = \qlam_1(t)C_{11}\dev{\qs}_1(\qr)+\qlam_1(t)C_{21}\dev{\qs}_2(\qr)+\qlam_1(t)C_{31}\dev{\qs}_3(\qr)\mathrel{+}\\
 &+\qlam_2(t)C_{12}\dev{\qs}_1(\qr)+\qlam_2(t)C_{22}\dev{\qs}_2(\qr)+\qlam_2(t)C_{32}\dev{\qs}_3(\qr)\mathrel{+}\\
 &+\qlam_3(t)C_{13}\dev{\qs}_1(\qr)+\qlam_3(t)C_{23}\dev{\qs}_2(\qr)+\qlam_3(t)C_{33}\dev{\qs}_3(\qr)\mathrel{+}\\
 &+\qlam_1(t)C_{11}\sph{\qs}_1(\qr)+\qlam_1(t)C_{21}\sph{\qs}_2(\qr)+\qlam_1(t)C_{31}\sph{\qs}_3(\qr)\mathrel{+}\\
 &+\qlam_2(t)C_{12}\sph{\qs}_1(\qr)+\qlam_2(t)C_{22}\sph{\qs}_2(\qr)+\qlam_2(t)C_{32}\sph{\qs}_3(\qr)\mathrel{+}\\
 &+\qlam_3(t)C_{13}\sph{\qs}_1(\qr)+\qlam_3(t)C_{23}\sph{\qs}_2(\qr)+\qlam_3(t)C_{33}\sph{\qs}_3(\qr),
 \end{split}\\  \label{zetrug2}
 \begin{split}
\rhe{\hhzet} \71{t, \qr} & = \qkap_1(t)\f{C_{11}}{\qeta_1}\dev{\qs}_1(\qr)+\qkap_1(t)\f{C_{21}}{\qeta_1}\dev{\qs}_2(\qr)+\qkap_1(t)\f{C_{31}}{\qeta_1}\dev{\qs}_3(\qr)\mathrel{+}\\
 &+\qkap_2(t)\f{C_{12}}{\qeta_2}\dev{\qs}_1(\qr)+\qkap_2(t)\f{C_{22}}{\qeta_2}\dev{\qs}_2(\qr)+\qkap_2(t)\f{C_{32}}{\qeta_2}\dev{\qs}_3(\qr)\mathrel{+}\\
 &+\qkap_3(t)\f{C_{13}}{\qeta_3}\dev{\qs}_1(\qr)+\qkap_3(t)\f{C_{23}}{\qeta_3}\dev{\qs}_2(\qr)+\qkap_3(t)\f{C_{33}}{\qeta_3}\dev{\qs}_3(\qr)\mathrel{+}\\
 &+\qkap_1(t)C_{11}\sph{\qs}_1(\qr)+\qkap_1(t)C_{21}\sph{\qs}_2(\qr)+\qkap_1(t)C_{31}\sph{\qs}_3(\qr)\mathrel{+}\\
 &+\qkap_2(t)C_{12}\sph{\qs}_1(\qr)+\qkap_2(t)C_{22}\sph{\qs}_2(\qr)+\qkap_2(t)C_{32}\sph{\qs}_3(\qr)\mathrel{+}\\
 &+\qkap_3(t)C_{13}\sph{\qs}_1(\qr)+\qkap_3(t)C_{23}\sph{\qs}_2(\qr)+\qkap_3(t)C_{33}\sph{\qs}_3(\qr),
 \end{split}
 \end{align}
and the rheological operators generate the equations
 \begin{align}
 \dev{\oS}\Big[\qlam_1(t)C_{11}+\qlam_2(t)C_{12}+\qlam_3(t)C_{13}\Big]&=\dev{\oZ}\left[\qkap_1(t)\f{C_{11}}{\qeta_1}+\qkap_2(t)\f{C_{12}}{\qeta_2}+\qkap_3(t)\f{C_{13}}{\qeta_3}\right],\\
 \dev{\oS}\Big[\qlam_1(t)C_{21}+\qlam_2(t)C_{22}+\qlam_3(t)C_{23}\Big]&=\dev{\oZ}\left[\qkap_1(t)\f{C_{21}}{\qeta_1}+\qkap_2(t)\f{C_{22}}{\qeta_2}+\qkap_3(t)\f{C_{23}}{\qeta_3}\right],\\
 \dev{\oS}\Big[\qlam_1(t)C_{31}+\qlam_2(t)C_{32}+\qlam_3(t)C_{33}\Big]&=\dev{\oZ}\left[\qkap_1(t)\f{C_{31}}{\qeta_1}+\qkap_2(t)\f{C_{32}}{\qeta_2}+\qkap_3(t)\f{C_{33}}{\qeta_3}\right],
 \end{align}
 \begin{align}
 \nonumber&\sph{\oS}\Big\{\qlam_1(t)\left[C_{11}+C_{21}\right]+\qlam_2(t)\left[C_{12}+C_{22}\right]+\qlam_3(t)\left[C_{13}+C_{23}\right]\Big\}=\\
 &\qquad\qquad\qquad=\sph{\oZ}\Big\{\qkap_1(t)\left[C_{11}+C_{21}\right]+\qkap_2(t)\left[C_{12}+C_{22}\right]+\qkap_3(t)\left[C_{13}+C_{23}\right]\Big\},
 \\
 \nonumber&\sph{\oS}\Big\{\qlam_1(t)\left[3C_{21}+C_{31}\right]+\qlam_2(t)\left[3C_{22}+C_{32}\right]+\qlam_3(t)\left[3C_{23}+C_{33}\right]\Big\}=\\
 &\qquad\qquad\qquad=\sph{\oZ}\Big\{\qkap_1(t)\left[3C_{21}+C_{31}\right]+\qkap_2(t)\left[3C_{22}+C_{32}\right]+\qkap_3(t)\left[3C_{23}+C_{33}\right]\Big\}
 ;
 \end{align}
while the boundary condition imposes
 \begin{align}
 \qlam(t)=\qlam_1(t)+\qlam_2(t)+\qlam_3(t).
 \end{align}

\subsubsection*{Free lateral deformations (\m{k=0})}

In this case, the complementary stress field can be represented again
as the sum of three independent spatial patterns: 
 \begin{align}
\el\hhsig(\qr)&=c_1 \81{\qeta} \qs_1(\qr)+c_2 \81{\qeta} \qs_2(\qr)+
c_3 \81{\qeta} \qs_3(\qr) ,
 \end{align}
 where 
 \begin{align}
 &&
c_1(\qeta) & = 1, & c_2(\qeta) & = \f{1-\qeta}{2+\qeta}, &
c_3(\qeta) & = \f{1-\qeta}{1+2\qeta} ,
 &&
 \end{align}
 and
 \begin{align}
 \nonumber
 s_{1,rr}(\qr)&=\f{\gamma R}{4}\92{-3\f{R}{r}\sin{\qphi}\biT+\biT\91{5\f{R^3}{r^3}-4\f{R^5}{r^5}}\sin{3\qphi}}\biT+\biT\f{\gamma d}{2}\92{\f{R^2}{r^2}\biT-\biT \91{4\f{R^2}{r^2}+3\f{R^4}{r^4}}\cos{2\qphi}},\\
 \nonumber
 s_{1,\qphi\qphi}(\qr)&=\f{\gamma R}{4}\left[\f{R}{r}\sin{\qphi}-\left(\f{R^3}{r^3}-4\f{R^5}{r^5}\right)\sin{3\qphi}\right]-\f{\gamma d}{2} \left(\f{R^2}{r^2}+3\f{R^4}{r^4}\cos{2\qphi}\right),\\
 \nonumber
 s_{1,r\qphi}(\qr)&=\f{\gamma R}{4}\left[-\f{R}{r}\cos{\qphi}-\left(3\f{R^3}{r^3}-4\f{R^5}{r^5}\right)\cos{3\qphi}\right]-\f{\gamma d}{2} \left(2\f{R^2}{r^2}-3\f{R^4}{r^4}\right)\sin{2\qphi},\\
 \label{sfree}
 s_{1,zz}(\qr)&=s_{1,rz}(\qr)=s_{1,\qphi z}(\qr)=0,\\
 \nonumber
\qs_2(\qr)&=\begin{pmatrix}
0&0&0
 \\[1ex]
0&0&0
 \\[1ex]
0&0&\gamma R\f{R^3}{r^3}\sin{3\qphi}-2\gamma d \f{R^2}{r^2}\cos{2\qphi}
 \end{pmatrix},\\
 \nonumber
\qs_3(\qr)&=\begin{pmatrix}
\f{\gamma R}{4}\left(-\f{R}{r}+\f{R^3}{r^3}\right)\sin{\qphi} &
\f{\gamma R}{4}\left(\f{R}{r}-\f{R^3}{r^3}\right)\cos{\qphi} &
0
 \\[1.5ex]
\f{\gamma R}{4}\left(\f{R}{r}-\f{R^3}{r^3}\right)\cos{\qphi} &
-\f{\gamma R}{4}\left(\f{R}{r}+\f{R^3}{r^3}\right)\sin{\qphi} &
0 
 \\[1.5ex]
0 & 0 & -\f{\gamma R}{2}\f{R}{r}\sin{\qphi}
 \end{pmatrix}.
 \end{align}
 The matrix \m{C_{jk}} is
 \begin{align}
 C_{jk}=\begin{pmatrix}
1 & 1 & 1
 \\[2.ex]
\displaystyle{\f{1-\qeta_1}{2+\qeta_1}} &
\displaystyle{\f{1-\qeta_2}{2+\qeta_2}} &
\displaystyle{\f{1-\qeta_3}{2+\qeta_3}}
 \\[2.ex]
\displaystyle{\f{1-\qeta_1}{1+2\qeta_1}} &
\displaystyle{\f{1-\qeta_2}{1+2\qeta_2}} &
\displaystyle{\f{1-\qeta_3}{1+2\qeta_3}}
 \end{pmatrix}.
 \end{align}
Now noticing that \m{\sph{\qs}_3=2\left(\sph{\qs}_1-\sph{\qs}_2\right)},
there is again one spherical equation less than deviatoric:
 \begin{align}
 \begin{split}
 \dev{\oS}\Big[\qlam_1(t)C_{11}+\qlam_2(t)C_{12}+\qlam_3(t)C_{13}\Big]&=\dev{\oZ}\left[\qkap_1(t)\f{C_{11}}{\qeta_1}+\qkap_2(t)\f{C_{12}}{\qeta_2}+\qkap_3(t)\f{C_{13}}{\qeta_3}\right],\\
 \dev{\oS}\Big[\qlam_1(t)C_{21}+\qlam_2(t)C_{22}+\qlam_3(t)C_{23}\Big]&=\dev{\oZ}\left[\qkap_1(t)\f{C_{21}}{\qeta_1}+\qkap_2(t)\f{C_{22}}{\qeta_2}+\qkap_3(t)\f{C_{23}}{\qeta_3}\right],\\
 \dev{\oS}\Big[\qlam_1(t)C_{31}+\qlam_2(t)C_{32}+\qlam_3(t)C_{33}\Big]&=\dev{\oZ}\left[\qkap_1(t)\f{C_{31}}{\qeta_1}+\qkap_2(t)\f{C_{32}}{\qeta_2}+\qkap_3(t)\f{C_{33}}{\qeta_3}\right],
 \end{split}
 \end{align}
 \begin{align}
 \begin{split}
& \sph{\oS}\Big\{\qlam_1(t)\left[C_{11}+2C_{31}\right]+\qlam_2(t)\left[C_{12}+2C_{32}\right]+\qlam_3(t)\left[C_{13}+2C_{33}\right]\Big\}=\\
 & \qquad\qquad\qquad
=\sph{\oZ}\Big\{\qkap_1(t)\left[C_{11}+2C_{31}\right]+\qkap_2(t)\left[C_{12}+2C_{32}\right]+\qkap_3(t)\left[C_{13}+2C_{33}\right]\Big\},\\
& \sph{\oS}\left\{\qlam_1(t)\left[C_{21}-2C_{31}\right]+\qlam_2(t)\left[C_{22}-2C_{32}\right]+\qlam_3(t)\left[C_{23}-2C_{33}\right]\right\}=\\
& \qquad\qquad\qquad
=\sph{\oZ} \33{ \qkap_1(t)\left[C_{21}-2C_{31}\right]+\qkap_2(t)\left[C_{22}-2C_{32}\right]+\qkap_3(t)\left[C_{23}-2C_{33}\right]
} .
 \end{split}
 \end{align}
Finally, from the boundary condition, we have
 \begin{align}
 \qlam(t)=\qlam_1(t)+\qlam_2(t)+\qlam_3(t).
 \end{align}

\subsubsection{Pressurizing of a thick-walled tube and a spherical tank?}

The elastic solution for these examples can be written in the form (see
Subsection~\ref{tubetank}):
 \begin{align}
\el\hhsig(\qr)&=c \81{\qeta} \qs(\qr),
 \\
\el\hhzet(\qr)&=\f{1}{\qeta}\dev{\qs}(\qr)+\sph{\qs}(\qr) ,
 \end{align}
where \m{c(\qeta)=1} and \m{\qs(\qr)=\el\hhsig(\qr)}. The elastic solution
of the problem with time dependent boundary condition is
 \begin{align}
\elt\hhsig \71{t, \qr} & = \qlam(t)\left[\dev{\qs}(\qr)+\sph{\qs}(\qr)\right],\\
\elt\hhzet \71{t, \qr} & = \qlam(t)\left[\f{1}{\qeta}\dev{\qs}(\qr)+\sph{\qs}(\qr)\right].
 \end{align}
Considering the rheological solution via this second method, our ansatz
is
 \begin{align}
\rhe\hhsig \71{t, \qr} & = \qlam(t)\left[\dev{\qs}(\qr)+\sph{\qs}(\qr)\right],\\
\rhe\hhzet \71{t, \qr} & = \kappa(t)\left[\f{1}{\qeta}\dev{\qs}(\qr)+\sph{\qs}(\qr)\right].
 \end{align}
The rheological operators generate the set of equations
 \begin{align}
 \dev{\oS}\qlam(t)=\dev{\oZ}\f{\kappa(t)}{\qeta},
 \qquad\qquad\qquad
 \sph{\oS}\qlam(t)=\sph{\oZ}\kappa(t),
 \end{align}
which is an overdetermined system of equations [two equations for the
only unknown \m{\kappa(t)}] so in this case the second method cannot
provide the solution. (Fortunately, the third method coming soon will
cover this case as well.)

\subsection{Conclusions about the first and second approaches}

We have seen that the first method is rather limited since only
those problems can be solved via the method
where the number of linearly independent spatial pattern deviatoric
parts, plus the number of linearly independent spatial pattern spherical
parts, is not more than three.
 
In the second approach there is no upper bound on the number of
independent spatial patterns. However, a limitation is that
only the freedom in \emph{stress} elastic patterns is utilized,
(stress-dimensioned) \emph{strain} is not considered, and the
independent deviatoric and spherical equations are not controlled,
either. Therefore, such simple problems as the pressurizing of
thick-walled tubes or of spherical tanks cannot be solved via the second
approach.
 
These experiences have inspired us to establish a common generalization
of the first two methods.

\subsection{Common generalization: The third approach: method of four
elastic spatial pattern sets} 
\label{third}

Since the second method prescribes separate equations related to the
deviatoric and spherical parts of the spatial patterns
\m{\qs_j \01{\qr}}, now the idea is to
explore the linear independence content within the deviatoric sector and
within the spherical one separately. Moreover, we reveal the linear
independence content not only for stress but also for stress-dimensioned
strain.
 
We assume again that the elastic stress solution can be written in a
finite sum form:
 \begin{align}  \label{sig3}
\el\hhsig\71{\qeta,\qr}=\sum_{i=1}^I \qa_i
\81{\qeta}\qalpha_i\01{\qr}=\sum_{i=1}^I \03{\qa_i
\01{\qeta}\02{\dev{\qalpha}_i\01{\qr}+\sph{\qalpha}_i\01{\qr}}},
 \end{align}
where \m{I} is some integer, \m{\qa_i(\qeta)}'s are linearly
independent coefficient functions and \m{\qalpha_i(\qr)}'s are linearly
independent spatial patterns (as before).
However, now we also
express the elastic stress-dimensioned strain solution in an
analogous finite sum:
 \begin{align}  \label{zeta3}
\el\hhzet\71{\qeta,\qr}=\sum_{j=1}^J \qb_j \81{\qeta}\qbeta_j\01{\qr}=
\sum_{j=1}^J \93{\qb_j \81{\qeta}\02{\dev{\qbeta}_j\01{\qr}+
\sph{\qbeta}_j\01{\qr}}},
 \end{align}
where \m{J} is a separate integer, \m{\qb_j(\qeta)}'s are
linearly independent coefficient functions [not necessarily the same
functions as the \m{\qa_i(\qeta)}'s] and \m{\qbeta_j(\qr)}'s are
linearly independent spatial patterns [not necessarily the same patterns
as the \m{\qalpha_i(\qr)}'s]. Finally, we write the deviatoric and
spherical parts of the elastic stress solution also as separate
expansions
 \begin{align}  \label{devsig3}
\el\hhsig^{\rm d}\71{\qeta,\qr}&=\sum_{k=1}^K \qeta \qc_k
\81{\qeta}\qgamma_k\01{\qr},\\
 \label{sphsig3}
\el\hhsig^{\rm s}\71{\qeta,\qr}&=\sum_{l=1}^L \qd_l
\81{\qeta}\qdelta_l\01{\qr} ,
 \end{align}
where \m{K} and \m{L} are integers, \m{\qc_k(\qeta)}'s and
\m{\qd_l(\qeta)}'s are linearly independent coefficient function sets,
and \m{\qgamma_k(\qr)}'s and \m{\qdelta_l(\qr)}'s are linearly
independent spatial pattern sets.

Based on Hooke's law in the form \re{delgamzet},
it is straightforward to find the conditions \m{K\le I}, \m{ L\le J
}, \m{I\le K+L}, \m{J\le K+L} for \m{I,\ J,\ K} and \m{L}.

When the boundary condition is multiplied by the time dependent factor
then the elastic stress solution is
 \begin{align}  \label{elt}
\elt\hhsig\71{t, \qeta, \qr}=\qlam\81{t}\el\hhsig \71{\qeta,\qr}.
 \end{align}
 
As in the first two methods, we wish to obtain the solution of the
rheological problem as (time dependent) combination of elastic
solutions, hence fulfilling the spatial condition \re{impp},
\re{compzeta} and the stress boundary condition automatically.
Let us fix \m{I} values
\m{\qeta_1,\qeta_2,\ldots,\qeta_m,\ldots,\qeta_I} for the stress
solution and \m{J} values
\m{\qeta_1,\qeta_2,\ldots,\qeta_n,\ldots,\qeta_J} for the
stress-dimensioned strain solution, and consider time dependent linear
combinations of the corresponding elastic solutions, separate ones
for stress and for the stress-dimensioned
strain (as before).

In light of \re{sig3} and \re{zeta3} , our ansatz is now
 \begin{align}  \label{sigreol3}
 \rhe{\hhsig}\71{t, \qr}&=\sum_{m=1}^I \qphi_m\81{t}\el\hhsig\71{\qeta_m,\qr}=
\sum_{m,i=1}^I \qphi_m\81{t} \qa_i \81{\qeta_m}\qalpha_i\01{\qr},\\
 \label{zetreol3}
 \rhe{\hhzet}\71{t, \qr}&=\sum_{n=1}^J \qpsi_n\81{t}\el\hhzet\71{\qeta_n,\qr}=
\sum_{n,j=1}^J \qpsi_n\81{t} \qb_j \81{\qeta_n}\qbeta_j\01{\qr}.
 \end{align}
Linear relationships hold among the deviatoric and spherical parts of
\m{\qalpha\01{\qr}}, \m{\qbeta\01{\qr}} and \m{\qgamma\01{\qr}},
\m{\qdelta\01{\qr}}, [see \re{sig3}--\re{delgamzet}]: with appropriate
matrices \m{\qA_{ik}}, \m{\qB_{jk}}, \m{\qC_{il}} and \m{\qD_{jl}},
 \begin{align}  \label{matrix1}
&&
\dev{\qalpha}_i\01{\qr}&=\sum_{k=1}^K\qA_{ik}\qgamma_k\01{\qr},&
\dev{\qbeta}_j\01{\qr}&=\sum_{k=1}^K\qB_{jk}\qgamma_k\01{\qr},
&&
 \\  \label{matrix2}
&&
\sph{\qalpha}_i\01{\qr}&=\sum_{l=1}^L\qC_{il}\qdelta_l\01{\qr},&
\sph{\qbeta}_j\01{\qr}&=\sum_{l=1}^L\qD_{jl}\qdelta_l\01{\qr}.
&&
 \end{align}
Now substituting \re{matrix1} and \re{matrix2} in \re{sigreol3} and
\re{zetreol3}, in the deviatoric--spherical separation one finds
 \begin{align}
 \rhe{\hhsig}^{\rm d}\71{t, \qr}&=\sum_{m,i=1}^I \92{\qphi_m\81{t}
\qa_i \81{\qeta_m}\sum_{k=1}^K\qA_{ik}\qgamma_k\01{\qr}},\\
 \rhe{\hhzet}^{\rm d}\71{t, \qr}&=\sum_{n,j=1}^J \92{\qpsi_n\81{t}
 \qb_j \81{\qeta_n}\sum_{k=1}^K \qB_{jk}\qgamma_k\01{\qr}},\\
 \rhe{\hhsig}^{\rm s}\71{t, \qr}&=\sum_{m,i=1}^I \92{\qphi_m\81{t}
\qa_i \81{\qeta_m}\sum_{l=1}^L\qC_{il}\qdelta_l\01{\qr}},\\
 \rhe{\hhzet}^{\rm s}\71{t, \qr}&=\sum_{n,j=1}^J \92{\qpsi_n\91{t}
\qb_j \91{\qeta_n}\sum_{l=1}^L\qD_{jl}\qdelta_l\01{\qr}}.
 \end{align}
The rheological conditions generate the system of equations
 \begin{align}
\dev{\oS}\sum_{m,i=1}^I \sum_{k=1}^K\qphi_m\81{t} \qa_i
\81{\qeta_m}\qA_{ik}\qgamma_k\01{\qr}&=
\dev{\oZ}\sum_{n,j=1}^J \sum_{k=1}^K\qpsi_n\81{t} \qb_j
\81{\qeta_n} \qB_{jk}\qgamma_k\01{\qr},
 \quad
&k&=1,\ldots,K\\
\sph{\oS}\sum_{m,i=1}^I \sum_{l=1}^L\qphi_m\81{t} \qa_i
\81{\qeta_m}\qC_{il}\qdelta_l\01{\qr}&=
\sph{\oZ}\sum_{n,j=1}^J \sum_{l=1}^L\qpsi_n\81{t} \qb_j
\81{\qeta_n}\qD_{jl}\qdelta_l\01{\qr},
 \quad
&l&=1,\ldots,L.
 \end{align}
Since the spatial patterns \m{\qdelta\01{\qr}}'s and \m{ \qgamma\01{\qr}
}'s are linearly independent sets, the equality of the corresponding
coefficients follows:
 \begin{align}
 \label{soldev3}
 \sum_{m,i=1}^I\dev{\oS}\qphi_m\81{t} \qa_i \81{\qeta_m}\qA_{ik}&=
 \sum_{n,j=1}^J\dev{\oZ}\qpsi_n\81{t} \qb_j \81{\qeta_n} \qB_{jk},
 \quad
 &k&=1,\ldots,K\\
 \label{solsph3}
 \sum_{m,i=1}^I\sph{\oS}\qphi_m\81{t} \qa_i \81{\qeta_m}\qC_{il}&=
 \sum_{n,j=1}^J\sph{\oZ}\qpsi_n\81{t} \qb_j \81{\qeta_n}\qD_{jl},
 \quad
 &l&=1,\ldots,L.
 \end{align}
The boundary condition requires
 \begin{align}
 \label{bc3}
 \sum_{m=1}^I \qphi_m\01{t}=\qlam\01{t} .
 \end{align}
Altogether, we have \m{K+L+1} equations for the \m{I+J} unknowns.

This is the
 most general version of our approach.
Compared to the first two methods, this third covers both, and it
extends the idea probably to the most general level reachable. Namely,
while the first two methods utilize the available elastic spatial
patterns in two restricted forms, the third formulation
explores
all
freedom
enabled by the stress spatial pattern set, the
stress-dimensioned strain one, the deviatoric pattern set and the
spherical one each. Again, let us next see via examples how to use it.

\subsubsection{Cylindrical bore (tunnel) and spherical hollow opened in
infinite, homogeneous and isotropic stress field}

As we have seen above, these two problems lead to the same rheological
equations so we treat them together again. The elastic stress
solution of the cylindrical bore case
can be written in the form of \re{sig3} as
 \begin{align}
\el\hhsig \01{\qr}=\bsig\begin{pmatrix}-\f{R^2}{r^2}&0&0\\0&\f{R^2}{r^2}&0\\0&0&0\end{pmatrix}=\sum_{i=1}^I
\qa_i \81{\qeta}\qalpha_i\01{\qr}
 \end{align}
with \m{I=1}, \m{\qa_1\01{\qeta}=1} and
\m{\qalpha_1\01{\qr}=\el\hhsig \01{\qr}}.
In parallel, the elastic stress solution of the spherical hollow case
is
 \begin{align}
\el\hhsig (\qr)=\bsig_{rr}\f{R^3}{r^3}\begin{pmatrix}-1&0&0\\
0&\f{1}{2}&0\\ 0&0&\f{1}{2}\end{pmatrix}=\sum_{i=1}^I \qa_i
\81{\qeta}\qalpha_i\01{\qr}
 \end{align}
with \m{I=1}, \m{\qa_1\01{\qeta}=1} and
\m{\qalpha_1\01{\qr}=\el\hhsig \01{\qr}} again. The deviatoric part of the
elastic stress in the form of \re{devsig3} is
 \begin{align}
\el\hhsig^{\rm d}\01{\qr}=\sum_{k=1}^K \qeta \qc_k
\81{\qeta}\qgamma_k\01{\qr}
 \end{align}
with \m{K=1}, \m{\qc_1\01{\qeta}=\f{1}{\qeta}} and
\m{\qgamma_1\01{\qr}=\el\hhsig \01{\qr}}. Since the spherical part of
the elastic stress is zero, \m{\el\hhsig^{\rm s}\01{\qr}=\zero},
which means that \m{L=0}. Similarly, with these, the stress-dimensioned
strain in the form of \re{zeta3} is
 \begin{align}
\el\hhzet \71{\qeta,\qr}&=\f{\sph{\qE}}{\dev{\qE}}\el\hhsig \01{\qr}=
\f{1}{\qeta}\el\hhsig \01{\qr}=\sum_{j=1}^J \qb_j \81{\qeta}\qbeta_j\01{\qr}
 \end{align}
with \m{J=1}, \m{\qb_1\01{\qeta}=\f{1}{\qeta}} and
\m{\qbeta_1\01{\qr}=\el\hhsig \01{\qr}}. The matrices \m{\qA_{ik}} and
\m{\qB_{jk}} can be read off easily, are \m{1\times 1} matrices,
namely, \m{\qA_{ik}=\qB_{jk}=1}. Therefore, subsituting into \re{soldev3},
\re{solsph3} and \re{bc3} provides the system of equations
 \begin{align}
 \dev{\oS}\qphi\01{t}&=\dev{\oZ}\qpsi\01{t}\f{1}{\qeta},\\
 \qphi\01{t}&=\qlam\01{t},
 \end{align}
two equations for the two unknown functions \m{\qphi\01{t}},
\m{\qpsi\01{t}}.

\subsubsection{Pressurizing of a thick-walled tube and a spherical tank}

The elastic stress solution of the thick-walled tube in the form of
\re{sig3} is
 \begin{align}
\el\hhsig \01{\qr}=p_0\f{R_i^2}{\RO^2-\RI^2}\begin{pmatrix}
1-\f{\RO^2}{r^2}&0&0\\0&1+\f{\RO^2}{r^2}&0\\0&0&0\end{pmatrix}
=\sum_{i=1}^I \qa_i \81{\qeta}\qalpha_i\01{\qr}
 \end{align}
with \m{I=1}, \m{\qa_1\01{\qeta}=1}, \m{\qalpha_1\01{\qr}=\hhsig\01{\qr}}.
Similarly,
 for
the spherical tank,
 \begin{align}
\el\hhsig \01{\qr}=p_0\f{\RI^3}{\RO^3-\RI^3}\begin{pmatrix}1-\f{\RO^3}{r^3}&0&0\\0&1+\f{\RO^3}{2r^3}&0\\0&0&1+\f{\RO^3}{2r^3}\end{pmatrix}=\sum_{i=1}^I
\qa_i \81{\qeta}\qalpha_i\01{\qr}
 \end{align}
with \m{I=1}, \m{\qa_1\01{\qeta}=1} and
\m{\qalpha_1\01{\qr}=\el\hhsig \01{\qr}} again. The deviatoric parts
for both cases can then be written as
 \begin{align}
\el\hhsig^{\rm d}\01{\qr}=\dev{\qalpha}_1\01{\qr}=\sum_{k=1}^K \qeta
\qc_k \81{\qeta}\qgamma_k\01{\qr}
 \end{align}
with \m{K=1}, \m{\qc_1\01{\qeta}=\f{1}{\qeta}} and
\m{\qgamma_1\01{\qr}=\dev{\hhsig}\01{\qr}=\dev{\qalpha_1}\01{\qr}}.
Analogously, the spherical part is
 \begin{align}
\el\hhsig^{\rm s}\01{\qr}=\sph{\qalpha}_1\01{\qr}=\sum_{l=1}^L \qd_l
\81{\qeta}\qdelta_l\01{\qr}
 \end{align}
with \m{L=1}, \m{\qd_1\01{\qeta}=1} and
\m{\qdelta_1\01{\qr}=\sph{\hhsig}\01{\qr}=\sph{\qalpha}_1\01{\qr}}.
Finally, the stress-dimensioned strain in the form of \re{zeta3} is
 \begin{align}
 \el\hhzet \71{\qeta,\qr}&=\f{1}{\qeta}\el\hhsig^{\rm d}\01{\qr}+
\el\hhsig^{\rm s}\01{\qr}=\sum_{j=1}^J \qb_j \81{\qeta}\qbeta_j\01{\qr}
 \end{align}
with \m{J=2}, \m{\qb_1\01{\qeta}=\f{1}{\qeta}},
\m{\qb_2\01{\qeta}=1}, \m{\qbeta_1\01{\qr}=\dev{\hhsig}\01{\qr}} and
\m{\qbeta_2\01{\qr}=\sph{\hhsig}\01{\qr}}. Reading off the matrices
results in
 \begin{align}
 &&
\qA_{ik}&=1,&
\qB_{jk}&=\begin{pmatrix}1\\0\end{pmatrix},&
\qC_{il}&=1,&
\qD_{jl}&=\begin{pmatrix}0\\1\end{pmatrix}.
 &&
 \end{align}
Substituting these into \re{soldev3}, \re{solsph3} and \re{bc3}, we find
the system of equations
 \begin{align}
 \dev{\oS}\qphi_1\01{t}&=\dev{\oZ}\qpsi_1\01{t}\f{1}{\qeta},\\
 \sph{\oS}\qphi_1\01{t}&=\sph{\oZ}\qpsi_2\01{t},\\
 \qphi_1\01{t}&=\qlam\01{t}
 \end{align}
for the three unknowns \m{\qphi_1\01{t}}, \m{\qpsi_1\01{t}},
\m{\qpsi_2\01{t}}.

\subsubsection{Cylindrical bore (tunnel) opened in infinite, homogeneous
and anisotropic stress field}

The elastic stress solution of the problem is
 \begin{align}
 \el\hhsig \01{\qr}=\sum_{i=1}^I \qa_i \81{\qeta}\qalpha_i\01{\qr}=\f{1-\qeta}{2+\qeta}\qalpha_1\01{\qr}+1\cdot\qalpha_2\01{\qr},
 \end{align}
which means \m{I=2}, \m{\qa_1\01{\qeta}=\f{1-\qeta}{2+\qeta}} and
\m{\qa_2\01{\qeta}=1}; furthermore, from \re{anizosig},
\m{\qalpha_1\01{\qr}=\qs_2\01{\qr}} and
\m{\qalpha_2\01{\qr}=\qs_1\01{\qr}}. The deviatoric part is
 \begin{align}
 \el\hhsig^{\rm d}\01{\qr}=\sum_{k=1}^K \qeta \qc_k \81{\qeta}\qgamma_k\01{\qr}=\f{1-\qeta}{2+\qeta}\qgamma_1\01{\qr}+1\cdot\qgamma_2\01{\qr}
 \end{align}
so \m{K=2}, \m{\qc_1\01{\qeta}=\f{1-\qeta}{\91{2+\qeta}}\qeta},
\m{\qc_2\01{\qeta}=\f{1}{\qeta}},
\m{\qgamma_1\01{\qr}=\dev{\qalpha}_1\01{\qr}},
\m{\qgamma_2\01{\qr}=\dev{\qalpha}_2\01{\qr}}.
The spherical part is
 \begin{align}
 \el\hhsig^{\rm s}\01{\qr}=\sum_{l=1}^L \qd_l \81{\qeta}\qdelta_l\01{\qr}=\f{3}{2+\qeta}\qdelta_1\01{\qr}
 \end{align}
so \m{L=1}, \m{\qd_1\01{\qeta}=\f{3}{2+\qeta}},
\m{\qdelta_1\01{\qr}=\sph{\qalpha}_1\01{\qr}=\sph{\qalpha}_2\01{\qr}}.
The stress-dimensioned strain in the form of \re{zeta3} is
 \begin{align}
\el\hhzet \71{\qeta,\qr} & = \sum_{j=1}^J \qb_j \81{\qeta}\qbeta_j
\01{\qr} = \f{1-\qeta}{\91{2+\qeta}\qeta} \21{\qgamma_1
\01{\qr}+2\qgamma_2 \01{\qr}} + \f{3}{2+\qeta}
\21{\qgamma_2\01{\qr}+\qdelta_1\01{\qr}},
 \end{align}
thus \m{J=2}, \m{\qb_1\01{\qeta}=\f{1-\qeta}{\91{2+\qeta}}},
\m{\qb_2\01{\qeta}=\f{3}{2+\qeta}},
\m{\qbeta_1\01{\qr}=\qgamma_1\01{\qr}+2\qgamma_2\01{\qr}},
\m{\qbeta_2\01{\qr}=\qgamma_2\01{\qr}+\qdelta_1\01{\qr}}. We find
 \begin{align}
 \qA_{ik}&=\begin{pmatrix}1&0\\0&1\end{pmatrix},&
 \qB_{jk}&=\begin{pmatrix}1&2\\0&1\end{pmatrix},&
 \qC_{il}&=\begin{pmatrix}1\\1\end{pmatrix},&
 \qD_{jl}&=\begin{pmatrix}0\\1\end{pmatrix}.
 \end{align}
Finally, the deviatoric equations are
 \begin{align}
 \dev{\oS}\93{\qa_1\81{\qeta_1}\qphi_1\01{t}+\qa_1\81{\qeta_2}\qphi_2\01{t}}&=
 \dev{\oZ}\93{\qb_1\81{\qeta_1}\qpsi_1\01{t}+\qb_1\81{\qeta_2}\qpsi_2\01{t}},\\
 \dev{\oS}\93{\qa_2\81{\qeta_1}\qphi_1\01{t}+\qa_2\81{\qeta_2}\qphi_2\01{t}}&=
 \dev{\oZ}\93{\92{2\qb_1\01{\qeta_1}+\qb_2\81{\qeta_1}}\qpsi_1\01{t}+
\92{2\qb_1\01{\qeta_2}+\qb_2\01{\qeta_2}}\qpsi_2\01{t}},
 \end{align}
while the spherical equation is
 \begin{align}
 \sph{\oS}\93{\92{\qa_1\01{\qeta_1}+\qa_2\81{\qeta_1}}\qphi_1\01{t}+
\92{\qa_1\01{\qeta_2}+\qa_2\81{\qeta_2}}\qphi_1\01{t}}&=
 \sph{\oZ}\93{\qb_2\81{\qeta_1}\qpsi_1\01{t}+\qb_2\81{\qeta_2}\qpsi_2\01{t}} .
 \end{align}
The condition from the boundary condition is
 \begin{align}
 \qphi_1\01{t}+\qphi_2\01{t}&=\qlam\01{t}.
 \end{align}

\subsubsection{Cylindrical bore (tunnel) opened in homogeneous medium
loaded by its self weight}

\subsubsection*{Hydrostatic initial stress state (\m{k=1})}

The elastic stress solution can be written as
 \begin{align}
\el\hhsig \01{\qr}=\sum_{i=1}^I \qa_i \81{\qeta}\qalpha_i\01{\qr}=
1\cdot\qalpha_1\01{\qr}+\f{1-\qeta}{1+2\qeta}\qalpha_2\01{\qr}
 \end{align}
so \m{I=2}, \m{\qa_1\01{\qeta}=1} and
\m{\qa_2\01{\qeta}=\f{1-\qeta}{1+2\qeta}},
\m{\qalpha_1\01{\qr}=\qs_1\01{\qr}} and
\m{\qalpha_2\01{\qr}=\qs_2\01{\qr}}, with the notations of \re{sighidro}.
The deviatoric part is
 \begin{align}
\el\hhsig^{\rm d}\01{\qr}=\sum_{k=1}^K \qeta \qc_k
\81{\qeta}\qgamma_k\01{\qr}= \f{1}{\qeta}\cdot\qgamma_1\01{\qr}+
\f{1-\qeta}{\91{1+2\qeta}\qeta}\qgamma_2\01{\qr} ,
 \end{align}
thus \m{K=2}, \m{\qc_1\01{\qeta}=\f{1}{\qeta}},
\m{\qc_2\01{\qeta}=\f{1-\qeta}{\91{1+2\qeta}\qeta}},
\m{\qgamma_1\01{\qr}=\dev{\qalpha}_1\01{\qr}},
\m{\qgamma_2\01{\qr}=\dev{\qalpha}_2\01{\qr}}. The spherical part is
 \begin{align}
\el\hhsig^{\rm s}\01{\qr}=\sum_{l=1}^L \qd_l
\81{\qeta}\qdelta_l\01{\qr}=\f{3}{1+2\qeta}\qdelta_1\01{\qr}
 \end{align}
so \m{L=1},  \m{\qd_1\01{\qeta}=\f{3}{1+2\qeta}},
\m{\qdelta_1\01{\qr}=2\sph{\qalpha}_1\01{\qr}=\sph{\qalpha}_2\01{\qr}}.
Using these, the stress-dimensioned strain is
 \begin{align}
\el\hhzet \71{\qeta,\qr}&=\sum_{j=1}^J \qb_j 
\81{\qeta}\qbeta_j\01{\qr} =
\f{1}{\qeta} \21{\qgamma_1 \01{\qr}+2\qgamma_2\01{\qr}}+
\f{3}{1+2\qeta} \21{-\qgamma_2\01{\qr}+\qdelta_1\01{\qr}},
 \end{align}
which means \m{J=2}, \m{\qb_1\01{\qeta}=\f{1}{\qeta}},
\m{\qb_2\01{\qeta}=\f{3}{1+2\qeta}},
\m{\qbeta_1\01{\qr}=\qgamma_1\01{\qr}+2\qgamma_2\01{\qr}},
\m{\qbeta_2\01{\qr}=-\qgamma_2\01{\qr}+\qdelta_1\01{\qr}}. Reading off
the matrices yields
 \begin{align}
 \qA_{ik}&=\begin{pmatrix}1&0\\0&1\end{pmatrix},&
 \qB_{jk}&=\begin{pmatrix}1&2\\0&-1\end{pmatrix},&
 \qC_{il}&=\begin{pmatrix}\f{1}{2}\\1\end{pmatrix},&
 \qD_{jl}&=\begin{pmatrix}0\\1\end{pmatrix}.
 \end{align}
The deviatoric equations are
 \begin{align}
 \dev{\oS}\93{\qa_1\81{\qeta_1}\qphi_1\01{t}+\qa_1\81{\qeta_2}\qphi_2\01{t}}&=
 \dev{\oZ}\93{\qb_1\81{\qeta_1}\qpsi_1\01{t}+\qb_1\81{\qeta_2}\qpsi_2\01{t}},\\
 \dev{\oS}\93{\qa_2\81{\qeta_1}\qphi_1\01{t}+\qa_2\81{\qeta_2}\qphi_2\01{t}}&
=\dev{\oZ}\93{\92{\qb_1\01{\qeta_1}-\qb_2\81{\qeta_1}}\qpsi_1\01{t}+
\92{\qb_1\01{\qeta_2}-\qb_2\81{\qeta_2}}\qpsi_2\01{t}}
 \end{align}
while the spherical equation is
 \begin{align}
\sph{\oS}\93{\92{\f{1}{2}\qa_1\01{\qeta_1}+\qa_2\81{\qeta_1}}\qphi_1\01{t}
+\92{\f{1}{2}\qa_1\01{\qeta_2}+\qa_2\81{\qeta_2}}\qphi_1\01{t}}=
 &
\sph{\oZ}\93{\qb_2\81{\qeta_1}\qpsi_1\01{t}+\qb_2\81{\qeta_2}\qpsi_2
\01{t}}.
 \end{align}
The boundary condition requires
 \begin{align}
 \qphi_1\01{t}+\qphi_2\01{t}&=\qlam\01{t}.
 \end{align}

\subsubsection*{No lateral deformations
(\m{k=\f{\nu}{1-\nu}=\f{1-\qeta}{1+2\qeta}})}

The elastic stress solution can be written in the form of \re{sig3} as
 \begin{align}
 \el\hhsig \01{\qr}=\sum_{i=1}^I \qa_i \81{\qeta}\qalpha_i\01{\qr}=
 \f{1-\qeta}{1+2\qeta}\qalpha_1\01{\qr}+\f{2+\qeta}{1+2\qeta}\qalpha_2\01{\qr}
+\f{3\91{1-\qeta}\qeta}{\91{2+\qeta}\91{1+2\qeta}}\qalpha_3\01{\qr}
 \end{align}
so \m{I=3}, \m{\qa_1\01{\qeta}=\f{1-\qeta}{1+2\qeta}}, 
\m{\qa_2\01{\qeta}=\f{2+\qeta}{1+2\qeta}},
\m{\qa_3\01{\qeta}=\f{3\91{1-\qeta}\qeta}{\91{2+\qeta}\91{1+2\qeta}}},
\m{\qalpha_1\01{\qr}=\qs_1\01{\qr}}, \m{\qalpha_2\01{\qr}=\qs_2\01{\qr}}
and \m{\qalpha_3\01{\qr}=\qs_3\01{\qr}} from \re{signolat}. Its
deviatoric part is
 \begin{align}
 \el\hhsig^{\rm d}\01{\qr}=\sum_{k=1}^K \qeta \qc_k \81{\qeta}\qgamma_k\01{\qr}=\f{1-\qeta}{1+2\qeta}\qgamma_1\01{\qr}+\f{2+\qeta}{1+2\qeta}\qgamma_2\01{\qr}+
\f{3\91{1-\qeta}\qeta}{\91{2+\qeta}\91{1+2\qeta}}\qgamma_3\01{\qr}
 \end{align}
so \m{K=3}, \m{\qc_1\01{\qeta}=\f{1-\qeta}{\91{1+2\qeta}\qeta}},
\m{\qc_2\01{\qeta}=\f{2+\qeta}{\91{1+2\qeta}\qeta}}, 
\m{\qc_3\01{\qeta}=\f{3\91{1-\qeta}\qeta}{\91{2+\qeta}\91{1+2\qeta}\qeta}},
\m{\qgamma_1\01{\qr}=\dev{\qalpha}_1\01{\qr}},
\m{\qgamma_2\01{\qr}=\dev{\qalpha}_2\01{\qr}},
\m{\qgamma_3\01{\qr}=\dev{\qalpha}_3\01{\qr}}. The spherical part is
 \begin{align}
 \el\hhsig^{\rm s}\01{\qr}=\sum_{l=1}^L \qd_l \81{\qeta}\qdelta_l\01{\qr}=
\f{3}{1+2\qeta}\qdelta_1\01{\qr}+\91{\f{6}{2+\qeta}+\f{3}{1+2\qeta}}\qdelta_2\01{\qr} ,
 \end{align}
thus \m{L=2},  \m{\qd_1\01{\qeta}=\f{3}{1+2\qeta}},
\m{\qd_2\01{\qeta}=\f{6}{2+\qeta}+\f{3}{1+2\qeta}},
\m{\qdelta_1\01{\qr}=\sph{\qalpha}_1\01{\qr}},
\m{\qdelta_2\01{\qr}=\sph{\qalpha}_3\01{\qr}}. The stress-dimensioned
strain is
 \begin{align}
 \el\hhzet \71{\qeta,\qr}&=\sum_{j=1}^J \qb_j \81{\qeta}\qbeta_j\01{\qr}=
\f{1-\qeta}{\91{1+2\qeta}\qeta}
\21{\qgamma_1\01{\qr}-2\qgamma_3\01{\qr}-2\qdelta_1\01{\qr}-2\qdelta_2\01{\qr}}\mathrel{+}\\
 \nonumber
 & \quad +\f{2+\qeta}{\91{1+2\qeta}\qeta}\21{\qgamma_2\01{\qr}+\qgamma_3\01{\qr}+
\qdelta_1\01{\qr}+\qdelta_2\01{\qr}}+\f{6}{2+\qeta}\91{-\f{1}{2}
\qgamma_3\01{\qr}+\qdelta_2\01{\qr}}
 \end{align}
so \m{J=3}, \m{\qb_1\01{\qeta}=\f{1-\qeta}{\91{1+2\qeta}\qeta}},
\m{\qb_2\01{\qeta}=\f{2+\qeta}{\91{1+2\qeta}\qeta}},
\m{\qb_3\01{\qeta}=\f{6}{2+\qeta}},
\m{\qbeta_1\01{\qr}=\qgamma_1\01{\qr}-2\qgamma_3\01{\qr}-2\qdelta_1\01{\qr}-2\qdelta_2\01{\qr}},
\m{\qbeta_2\01{\qr}=\qgamma_2\01{\qr}+\qgamma_3\01{\qr}+\qdelta_1\01{\qr}+\qdelta_2\01{\qr}},
\m{\qbeta_3\01{\qr}=-\f{1}{2}\qgamma_3\01{\qr}+\qdelta_2\01{\qr}}. The
matrices are
 \begin{align}
 \qA_{ik}&=\begin{pmatrix}1&0&0\\0&1&0\\0&0&1\end{pmatrix},&
 \qB_{jk}&=\begin{pmatrix}1&0&-2\\0&1&1\\0&0&-\f{1}{2}\end{pmatrix},&
 \qC_{il}&=\begin{pmatrix}1&0\\1&3\\0&1\end{pmatrix},&
 \qD_{jl}&=\begin{pmatrix}-2&-2\\1&1\\0&1\end{pmatrix}.
 \end{align}
The rheological equations are
 \begin{align}
 &\dev{\oS}\93{\qa_1\81{\qeta_1}\qphi_1\01{t}+\qa_1\81{\qeta_2}\qphi_2\01{t}
+\qa_1\81{\qeta_3}\qphi_3\01{t}}=\\
 \nonumber
 &\qquad\dev{\oZ}\93{\qb_1\81{\qeta_1}\qpsi_1\01{t}+
\qb_1\81{\qeta_2}\qpsi_2\01{t}+\qb_1\81{\qeta_3}\qpsi_3\01{t}},\\
 &\dev{\oS}\93{\qa_2\81{\qeta_1}\qphi_1\01{t}+\qa_2\81{\qeta_2}\qphi_2\01{t}
+\qa_2\81{\qeta_3}\qphi_3\01{t}}=\\
 \nonumber
 &\qquad\dev{\oZ}\93{\qb_2\81{\qeta_1}\qpsi_1\01{t}+
\qb_2\81{\qeta_2}\qpsi_2\01{t}+\qb_2\81{\qeta_3}\qpsi_3\01{t}},\\
 &\dev{\oS}\93{\qa_3\81{\qeta_1}\qphi_1\01{t}+
\qa_3\81{\qeta_2}\qphi_2\01{t}+\qa_3\81{\qeta_3}\qphi_3\01{t}}=\\
 \nonumber
 &\qquad\dev{\oZ}\bigg\{\92{-2\qb_1\81{\qeta_1}+\qb_2\81{\qeta_1}-
\f{1}{2}\qb_3\81{\qeta_1}}\qpsi_1\01{t}+\92{-2\qb_1\81{\qeta_2}+
\qb_2\81{\qeta_2}-\f{1}{2}\qb_3\81{\qeta_2}}\qpsi_2\01{t} \mathrel+ \\
 \nonumber
 &\qquad\qquad+\92{-2\qb_1\81{\qeta_3}+\qb_2\81{\qeta_3}-
\f{1}{2}\qb_3\01{\qeta_3}}\qpsi_3\01{t}\bigg\},
 \end{align}
 \begin{align}
 &\sph{\oS}\93{\92{\qa_1\81{\qeta_1}+\qa_2\81{\qeta_1}}\qphi_1\01{t}+
\92{\qa_1\01{\qeta_2}+\qa_2\81{\qeta_2}}\qphi_1\01{t}
  +\92{\qa_1\01{\qeta_3}+\qa_2\81{\qeta_3}}\qphi_3\01{t}}=\\
 \nonumber
 &\qquad\sph{\oZ}\93{\92{-2\qb_1\01{\qeta_1}+\qb_2\81{\qeta_1}}\qpsi_1\01{t}
+\92{-2\qb_1\01{\qeta_2}+\qb_2\01{\qeta_2}}\qpsi_2\01{t}
+\92{-2\qb_1\01{\qeta_3}+\qb_2\01{\qeta_3}}\qpsi_3\01{t}},\\
 &\sph{\oS}\93{\92{3\qa_2\01{\qeta_1}+\qa_3\01{\qeta_1}}\qphi_1\01{t}
+\92{3\qa_1\01{\qeta_2}+\qa_3\01{\qeta_2}}\qphi_1\01{t}
+\92{3\qa_2\01{\qeta_3}+\qa_3\01{\qeta_3}}\qphi_3\01{t}}=\\
 \nonumber
 &\qquad\sph{\oZ}\big\{\92{-2\qb_1\01{\qeta_1}+\qb_2\01{\qeta_1}+
\qb_3\01{\qeta_1}}\qpsi_2\01{t}+
\92{-2\qb_1\01{\qeta_2}+\qb_2\01{\qeta_2}+\qb_3\01{\qeta_2}}\qpsi_2\01{t}
\mathrel+
 \\
 \nonumber
 &\qquad\qquad+\92{-2\qb_1\01{\qeta_3}+\qb_2\01{\qeta_3} +
\qb_3\01{\qeta_3}}\qpsi_3\01{t}\big\},
 \end{align}
and from the boundary condition we have
 \begin{align}
 \qphi_1\01{t}+\qphi_2\01{t}+\qphi_3\01{t}=\qlam\01{t}.
 \end{align}

\subsubsection*{Free lateral deformations (\m{k=0})}

The elastic stress solution can be written in the form of \re{sig3} as
 \begin{align}
 \el\hhsig \01{\qr}=\sum_{i=1}^I \qa_i \81{\qeta}\qalpha_i\01{\qr}=
 1\cdot\qalpha_1\01{\qr}+\f{1-\qeta}{2+\qeta}\qalpha_2\01{\qr}+\f{1-\qeta}{1+2\qeta}\qalpha_3\01{\qr},
 \end{align}
which means that \m{I=3}, \m{\qa_1\01{\qeta}=1}, 
\m{\qa_2\01{\qeta}=\f{1-\qeta}{2+\qeta}} and
\m{\qa_3\01{\qeta}=\f{1-\qeta}{1+2\qeta}},
\m{\qalpha_1\01{\qr}=\qs_1\01{\qr}}, \m{\qalpha_2\01{\qr}=\qs_2\01{\qr}}
and \m{\qalpha_3\01{\qr}=\qs_3\01{\qr}}, where \m{\qs_k\01{\qr}} is given
in \re{sfree}. The deviatoric part of the elastic stress is
 \begin{align}
\el\hhsig^{\rm d}\01{\qr}=\sum_{k=1}^K \qeta \qc_k \81{\qeta}\qgamma_k\01{\qr}=
1\cdot\qgamma_1\01{\qr}+\f{1-\qeta}{2+\qeta}\qgamma_2\01{\qr}+\f{1-\qeta}{1+2\qeta}\qgamma_3\01{\qr}
 \end{align}
so \m{K=3}, with \m{\qc_1\01{\qeta}=\f{1}{\qeta}},
\m{\qc_2\01{\qeta}=\f{1-\qeta}{\91{2+\qeta}\qeta}}, 
\m{\qc_3\01{\qeta}=\f{1-\qeta}{\91{1+2\qeta}\qeta}},
\m{\qgamma_1\01{\qr}=\dev{\qalpha}_1\01{\qr}},
\m{\qgamma_2\01{\qr}=\dev{\qalpha}_2\01{\qr}},
\m{\qgamma_3\01{\qr}=\dev{\qalpha}_3\01{\qr}}. The spherical part is
 \begin{align}
 \el\hhsig^{\rm s}\01{\qr}=\sum_{l=1}^L \qd_l \81{\qeta}\qdelta_l\01{\qr}=
\f{3}{1+2\qeta}\qdelta_1\01{\qr}+\f{3\qeta-3}{\91{2+\qeta}\91{1+2\qeta}}\qdelta_2\01{\qr} ,
 \end{align}
thus \m{L=2}, with \m{\qd_1\01{\qeta}=\f{3}{1+2\qeta}},
\m{\qd_2\01{\qeta}=\f{3\qeta-3}{\91{2+\qeta}\91{1+2\qeta}}},
\m{\qdelta_1\01{\qr}=\sph{\qalpha}_1\01{\qr}} and
\m{\qdelta_2\01{\qr}=\sph{\qalpha}_2\01{\qr}}. The stress-dimensioned
strain is
 \begin{align}
\el\hhzet \71{\qeta,\qr}&=\sum_{j=1}^J \qb_j \81{\qeta}\qbeta_j\01{\qr}=
\f{1}{\qeta}\21{\qgamma_1\01{\qr}+\qgamma_3\01{\qr}+\qdelta_2\01{\qr}}
 \nonumber \\
& \quad +\f{1-\qeta}{\91{2+\qeta}\qeta}\21{\qgamma_2\01{\qr}-2\qdelta_2\01{\qr}}+
\f{3}{1+2\qeta}\21{-\qgamma_3\01{\qr}+\qdelta_1\01{\qr}-\qdelta_2\01{\qr}}
 \end{align}
so \m{J=3}, \m{\qb_1\01{\qeta}=\f{1}{\qeta}},
\m{\qb_2\01{\qeta}=\f{1-\qeta}{\91{2+\qeta}\qeta}},
\m{\qb_3\01{\qeta}=\f{3}{1+2\qeta}},
\m{\qbeta_1\01{\qr}=\qgamma_1\01{\qr}+\qgamma_3\01{\qr}+\qdelta_2\01{\qr}},
\m{\qbeta_2\01{\qr}=\qgamma_2\01{\qr}-2\qdelta_2\01{\qr}},
\m{\qbeta_3\01{\qr}=-\qgamma_3\01{\qr}+\qdelta_1\01{\qr}-\qdelta_2\01{\qr}}.
One can read off the matrices as
 \begin{align}
 \qA_{ik}&=\begin{pmatrix}1&0&0\\0&1&0\\0&0&1\end{pmatrix},&
 \qB_{jk}&=\begin{pmatrix}1&0&1\\0&1&0\\1&0&-1\end{pmatrix},&
 \qC_{il}&=\begin{pmatrix}1&0\\0&1\\2&-2\end{pmatrix},&
 \qD_{jl}&=\begin{pmatrix}0&1\\0&-2\\1&-1\end{pmatrix}.
 \end{align}
With these, the deviatoric rheological equations are
 \begin{align}
 \nonumber
 &\dev{\oS}\93{\qa_1\81{\qeta_1}\qphi_1\01{t}+\qa_1\81{\qeta_2}\qphi_2\01{t}
+\qa_1\81{\qeta_3}\qphi_3\01{t}}=\\
 \nonumber
 &\qquad\dev{\oZ}\93{\qb_1\81{\qeta_1}\qpsi_1\01{t}+
\qb_1\81{\qeta_2}\qpsi_2\01{t}+\qb_1\81{\qeta_3}\qpsi_3\01{t}},\\
 &\dev{\oS}\93{\qa_2\81{\qeta_1}\qphi_1\01{t}+\qa_2\81{\qeta_2}\qphi_2\01{t}
+\qa_2\81{\qeta_3}\qphi_3\01{t}}=\\
 \nonumber
 &\qquad\dev{\oZ}\93{\qb_2\81{\qeta_1}\qpsi_1\01{t}+
\qb_2\81{\qeta_2}\qpsi_2\01{t}+\qb_2\81{\qeta_3}\qpsi_3\01{t}},\\
 \nonumber
 &\dev{\oS}\93{\qa_3\81{\qeta_1}\qphi_1\01{t}+\qa_3\81{\qeta_2}\qphi_2\01{t}
+\qa_3\81{\qeta_3}\qphi_3\01{t}}=\\
 \nonumber
 &\qquad\dev{\oZ}\93{\92{\qb_1\01{\qeta_1}-\qb_3\01{\qeta_1}}\qpsi_1\01{t}
+\92{\qb_1\01{\qeta_2}-\qb_3\01{\qeta_2}}\qpsi_2\01{t}
+\92{\qb_1\01{\qeta_3}-\qb_3\01{\qeta_3}}\qpsi_3\01{t}},
 \end{align}
and the spherical equations are
 \begin{align}
 \nonumber
 &\sph{\oS}\93{\92{\qa_1\01{\qeta_1}+2\qa_3\01{\qeta_1}}\qphi_1\01{t}
+\92{\qa_1\01{\qeta_2}+2\qa_3\01{\qeta_2}}\qphi_1\01{t}
+\92{\qa_1\01{\qeta_3}+2\qa_3\01{\qeta_3}}\qphi_3\01{t}}=\\
 \nonumber
 &\qquad\sph{\oZ}\93{\qb_3\81{\qeta_1}\qpsi_1\01{t}+
\qb_3\81{\qeta_2}\qpsi_2\01{t}+\qb_3\81{\qeta_3}\qpsi_3\01{t}},\\
 &\sph{\oS}\93{\92{\qa_2\01{\qeta_1}-2\qa_3\01{\qeta_1}}\qphi_1\01{t}
+\92{\qa_1\01{\qeta_2}-2\qa_3\01{\qeta_2}}\qphi_1\01{t}
+\92{\qa_2\01{\qeta_3}-2\qa_3\01{\qeta_3}}\qphi_3\01{t}}=\\
 \nonumber
 &\qquad\sph{\oZ}\{\92{\qb_1\01{\qeta_1}-2\qb_2\01{\qeta_1}-
\qb_3\81{\qeta_1}}\qpsi_1\01{t}+
 \92{\qb_1\01{\qeta_2}-2\qb_2\01{\qeta_2}-\qb_3\01{\qeta_2}}\qpsi_2\01{t}
\mathrel{+}\\
 \nonumber
 &\qquad\qquad+\92{\qb_1\01{\qeta_3}-2\qb_2\01{\qeta_3}-\qb_3\01{\qeta_3}}
\qpsi_3\01{t}\}.
 \end{align}
The boundary condition requires to fulfil
 \begin{align}
 \qphi_1\01{t}+\qphi_2\01{t}+\qphi_3\01{t}=\qlam\01{t}.
 \end{align}

\Section{Solutions of the previous examples with concrete material models}
\label{concrete}

All results of each of the three methods have so far been presented for
an arbitrary linear rheological model. In this section, we show the
solutions for certain concrete material models.

The function that scales the boundary condition is chosen as [\cf
\re{lambda}]:
 \begin{align}
\qlam(t) =
 \left\{  \begin{matrix}
0 & {\rm if} & t \le t_1 ,
 \\ 
{
\displaystyle
\f{1}{2} }
 \left[ 1 + \sin{ \left( \pi
{
\displaystyle
\f{t - \f{t_1+t_2}{2}}{t_2 - t_1}
}
\right) } \right]
& {\rm if} & t_1 \le t \le t_2,
\\ 1 & {\rm if} & t_2 \le t .
 \end{matrix}  \right.
 \end{align}

According to experience, rheological material behaviour is usually seen
in the deviatoric part, hence, in our examples we always assume Hooke's
elasticity model for the spherical part. In parallel, we take the Kelvin
model and the Kluitenberg--Verhás model for the deviatoric part.

Three cases are analyzed in all of the considered examples:
 \itemsep 0ex\parsep 0ex\topsep 0ex\partopsep -3ex\parskip 0ex
 \begin{itemize}
  \itemsep 0ex\parsep 0ex\topsep 0ex\partopsep -3ex\parskip 0ex
  \item when the switch-on \m{\qlam(t)} is very slow compared to the rheological time scales,
  \item when the switch-on time scale \m{t_2-t_1} is comparable to the rheological time scales,
  \item when the switch-on \m{\qlam(t)} is very fast compared to the rheological time scales.
 \end{itemize}
 
In the figures below, the functions
 \begin{align}
\Lambda_j^{\rm d}(t)=\sum_{k=1}^J\qlam_k(t) C_{jk},
 \qquad
{\rm K}_j^{\rm d}(t)=\sum_{k=1}^J\qkap_k(t) \f{1}{\qeta_k}C_{jk},
 \qquad\qquad
j&=1,\ldots,J
 \end{align}
indicate the time dependent factors of the deviatoric spatial patterns
of stress and of stress-dimensioned strain; and the analogous time
dependent coefficients of the spherical patterns are \m{\Lambda_j^{\rm
s}(t)} and \m{{\rm K}_j^{\rm s}(t)}; corresponding colour codes are
defined in \rf{legendd} (in cases when less functions are enough then
the rest are omitted).\footnote{Note that, while there is arbitrariness
in the choice of the constants \m { \qeta_k }, the spatial patterns are
fixed by the nature of the elastic solution so the coefficients of the
patterns do not contain any arbitrariness.}

 \begin{figure}[H]  \centering
   \ig[width=0.12\textwidth]{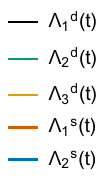}
  \hskip 0.3\textwidth
   \ig[width=0.12\textwidth]{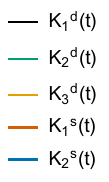}
\caption{\textsl{Left:} colour codes of the coefficient functions
belonging to the spatial patterns of stress. \textsl{Right:} colour
codes of the coefficient functions belonging to the spatial patterns of
stress-dimensioned strain.}
  \label{legendd}
 \end{figure}

Furthermore, in plots obtained by the third method, the notations
 \begin{align}  \label{hehe1}
 &&
\qPhid_k\01{t} & =
\sum_{m,i=1}^I\qphi_m\81{t}\qa_i \81{\qeta_m}\qA_{ik} ,
 &
\qPhis_l\01{t} & =
\sum_{m,i=1}^I\qphi_m\81{t}\qa_i \81{\qeta_m}\qC_{il} ,
 &&
 \\  \label{hehe2}
 &&
\qPsid_k\01{t} & =
\sum_{n,j=1}^J \qpsi_n\81{t} \qb_j \81{\qeta_n} \qB_{jk} ,
 &
\qPsis_l\01{t} & =
\sum_{n,j=1}^J \qpsi_n\81{t} \qb_j \81{\qeta_n} \qD_{jl}
 &&
 \end{align}
are used.
 
The following figures show the time evolution of the patterns belonging
to stress on the left the time evolution of the patterns belonging to
the stress-dimensioned strain is shown on the right. The cases of slow,
medium and fast changes in the boundary condition are displayed in the
order up-to-down.

\subsection{Kelvin -- Hooke model}

First consider the simplest rheological material model, Kelvin model in
the deviatoric part and Hooke model in the spherical part:
 \begin{align}
 &&
\dev{\qqsig}&=\qeta\dev{\qqzet}+\f{\qEd_1}{\qEs}\dev{\dot{\qqzet}},
 &
\sph{\qqsig}&=\sph{\qqzet}.
 &&
 \end{align}
The rheological time scale of the model is
\m{\f{\F{\qEd_1}{\qEs}}{\qeta}}. Using this as the unit of
time, the switch-on of the boundary condition is chosen in the slow
case as \m{t_2-t_1=5}, in the medium case as \m{t_2-t_1=1}, while in the
fast case as \m{t_2-t_1=0.1}.
All figures are calculated with \mm { \qeta = \F{\qEd}{\qEs} =
\F{\qEd_0}{\qEs} = 0.4 } [to which the Poisson's ratio \m { \nu =
0.25 } corresponds, \cf \re{nu}].
 
The simplest examples are the cylindrical bore (tunnel) and spherical
hollow opened in homogeneous and isotropic stress field, which lead to the
same deviatoric equation \re{80}, whose solution is plotted in
\rf{izo}. Due to the geometrical simplicity of the setting, we cannot
observe any remarkable phenomenon but it is to be noted that the
only coefficient function that belongs to stress is equal to the coefficient
function \m { \qlam(t) } of the boundary condition, and strain has an
observably slower increase so it can be easily recognized that
the medium gets deformed even after the end of the drilling.

As this problem is solvable via the first method, too, in \rf{izo-e} we
plot the corresponding time dependent function that replaces the static
deviatoric elasticity coefficient. As expected, the ratio \m { \qEd\1 1
{t} / \qEs } converges to the static---Hookean---value, \m { \qEd /
\qEs = \qeta = 0.4 }.

For the example of pressurizing of a thick-walled tube and of a
spherical tank, we find one deviatoric coefficient function and two
spherical ones (\rf{cso}). Correspondingly, in the first method, the
deviatoric and spherical elasticity constants are replaced with separate
time dependent functions (\rf{cso-e}). Here, the deviatoric ratio tends
again to the Hookean value \m { \qeta = 0.4 }, as it should.

In the case of the cylindrical bore (tunnel) opened in homogeneous but
anisotropic stress field, the situation is the opposite: One has two
deviatoric coefficient functions and a single spherical one, shown in
\rf{anizo}. In this case one can observe an interesting phenomenon: as
long as the boundary condition changes, there are strong transients in
the solution, and some of them are visible even after the end of the
change in the boundary condition. Also, this is a case where, according
to the first method, altogether three time dependent functions replace
the elasticity constants in stress and in strain (\rf{anizo-e}). The
two deviatoric ratios, \m { \qEds\1 1 {t} / \qEs } and \m { \qEde\1 1
{t} / \qEs }, tend to \m { \qeta = 0.4 } and the spherical one, \m {
\qEse\1 1 {t} / \qEs } to \m { 1 }, following the expectations.

Next, we present the solutions of the cylindrical bore (tunnel) opened
in homogeneous medium loaded by its self weight field. The solutions for
the hydrostatic initial stress state \m{(k=1)} can be seen in
\rf{hidro}. Here it is also observable that one of the coefficient
functions starts with anomalous sign---with opposite sign compared to
its eventual sign in the large-time asymptotics---. Consequently,
\textit{rheology means not only damping and delay but a more complicated
mechanical behaviour.}
 
When the primary field does not allow lateral deformations
\m{(k=\f{\nu}{1-\nu}=\f{1-\qeta}{1+2\qeta})} then there are three
independent deviatoric and two independent spherical patterns so we seek
for five-five functions; these can be seen in \rf{gatolt}.
 
Finally considered is the case of free lateral deformations \m{(k=0)}.
The solutions are plotted in \rf{szabad}. Time dependence is more
complicated here, three of the five functions starting with anomalous
sign.
 
The question arises what it may cause when the coefficient function of a
pattern starts with anomalous sign. We give the answer for this question
later, during the analysis of the displacement field.

\fffigg{0.475}{izotrop_alagut_s_lassu.pdf}
              {izotrop_alagut_s_koz.pdf}
              {izotrop_alagut_s_gyors.pdf}
              {izotrop_alagut_z_lassu.pdf}
              {izotrop_alagut_z_koz.pdf}
              {izotrop_alagut_z_gyors.pdf}
{Time evolution of the coefficient functions for the example of
a cylindrical bore (tunnel) opened in homogeneous and isotropic stress
field.
 \LeftRight{}{}  }{izo}

\fffig{0.44}{izo_e_lassu_color.pdf}
            {izo_e_koz_color.pdf}
            {izo_e_gyors_color.pdf}
{Time dependence at the place of the elasticity coefficient according to
the first method, for the example of a cylindrical bore (tunnel) opened
in homogeneous and isotropic stress field. In black, the gradual opening
 {\m { \qlam(t) }} is also displayed.
 \UpDown  }{izo-e}

\fffigg{0.475}{cso_s_lassu_color.pdf}
              {cso_s_koz_color.pdf}
              {cso_s_gyors_color.pdf}
              {cso_z_lassu_color.pdf}
              {cso_z_koz_color.pdf}
              {cso_z_gyors_color.pdf}
{Time evolution of the coefficient functions
for the example of pressurizing of a thick-walled tube
and of a spherical tank.
 \LeftRight{}{s}
The black denoted coefficients (that of stress and of spherical
stress-dimensioned strain) happen to coincide with \m { \qlam \1 1 {t}
}; green denotes the coefficient of deviatoric stress-dimensioned
strain.
 }{cso}

\fffig{0.44}{cso_e_lassu_color.pdf}
            {cso_e_koz_color.pdf}
            {cso_e_gyors_color.pdf}
{Time dependence at the place of the elasticity coefficients (orange:
deviatoric, red: spherical) according to the first method, for the
example of pressurizing of a thick-walled tube and of a spherical tank.
In black, the gradual opening \m { \qlam(t) } is also displayed.
 \UpDown  }{cso-e}

\fffigg{0.475}{anizotrop_alagut_s_lassu_color.pdf}
              {anizotrop_alagut_s_koz_color.pdf}
              {anizotrop_alagut_s_gyors_color.pdf}
              {anizotrop_alagut_z_lassu_color.pdf}
              {anizotrop_alagut_z_koz_color.pdf}
              {anizotrop_alagut_z_gyors_color.pdf}
{Time evolution of the coefficient functions for the example of a
cylindrical bore (tunnel) opened in homogeneous but anisotropic stress
field.
 \LeftRight{s}{s}  }{anizo}

\fffig{0.44}{anizo_e_lassu_color.pdf}
            {anizo_e_koz_color.pdf}
            {anizo_e_gyors_color.pdf}
{Time dependence at the place of the elasticity coefficients according
to the first method, for the example of a cylindrical bore (tunnel)
opened in homogeneous but anisotropic stress field. Black: the gradual
opening \m { \qlam }, green: \m{ \f{\qEds}{\qEs} }, orange: \m{
\f{\qEde}{\qEs} }, red: \m { \f{\qEse}{\qEs} }.
 \UpDown  }{anizo-e}

\fffigg{0.475}{onsuly_lam1_s_lassu_color.pdf}
              {onsuly_lam1_s_koz_color.pdf}
              {onsuly_lam1_s_gyors_color.pdf}
              {onsuly_lam1_z_lassu_color.pdf}
              {onsuly_lam1_z_koz_color.pdf}
              {onsuly_lam1_z_gyors_color.pdf}
{Time evolution of the coefficient functions for the example of a
cylindrical bore (tunnel) opened in homogeneous medium loaded by its
self weight, with hydrostatic initial stress state (\m{k=1}).
 \LeftRight{s}{s}  }{hidro}

\fffigg{0.475}{onsuly_lamnu_s_lassu_color.pdf}
              {onsuly_lamnu_s_koz_color.pdf}
              {onsuly_lamnu_s_gyors_color.pdf}
              {onsuly_lamnu_z_lassu_color.pdf}
              {onsuly_lamnu_z_koz_color.pdf}
              {onsuly_lamnu_z_gyors_color.pdf}
{Time evolution of the coefficient functions for the example of a
cylindrical bore (tunnel) opened in homogeneous medium loaded by its
self weight, with no lateral deformations allowed in the primary field
(\m{k=\f{\nu}{1-\nu}=\f{1-\qeta}{1+2\qeta}}).
 \LeftRight{s}{s}  }{gatolt}

\fffigg{0.475}{onsuly_lam0_s_lassu_color.pdf}
              {onsuly_lam0_s_koz_color.pdf}
              {onsuly_lam0_s_gyors_color.pdf}
              {onsuly_lam0_z_lassu_color.pdf}
              {onsuly_lam0_z_koz_color.pdf}
              {onsuly_lam0_z_gyors_color.pdf}
{Time evolution of the coefficient functions for the example of a
cylindrical bore (tunnel) opened in homogeneous medium loaded by its
self weight, with free lateral deformations in the primary field
(\m{k=0}).
 \LeftRight{s}{s}  }{szabad}

\subsection{Kluitenberg--Verhás -- Hooke model}

Next, we use the Kluitenberg--Verhás model in the deviatoric part. For
simplicity, we choose the coefficient of the time derivative of the
stress to zero---in this case the the index of inertia (see
\cite{asszonyi_fulop_van}) is necessarily positive, so (damped)
rheological oscillation will be present---:
 \begin{align}
 &&
\dev{\qqsig}&=\qeta\dev{\qqzet}+\f{\qEd_1}{\qEs}\dev{\dot{\qqzet}}+\f{\qEd_2}{\qEs}\dev{\ddot{\qqzet}},
 &
\sph{\qqsig}&=\sph{\qqzet}.
 &&
 \end{align}
 
Again, we use the rheological time scale
\m{\f{\F{\qEd_1}{\qEs}}{\qeta}} as time unit, and take \mm { \qeta
= 0.4 } [Poisson's ratio \mm{ \nu = 0.25 }].
 
The intensity of the oscillation is determined by the value of the
coefficient \m{\f{\qEd_2}{\qEs}}. We analyze two cases, when
\m{\f{\F{\qEd_2}{\qEs}}{\qeta}=0.1}, as well as when
\m{\f{\F{\qEd_2}{\qEs}}{\qeta}=1} (strongly and weakly damped
oscillation).
 
We consider the example of the cylindrical bore (tunnel) opened in
infinite, homogeneous but anisotropic stress field. In case of weak
oscillation the time evolution of the coefficient functions are plotted
in \rf{anizo_gyenge}, while in case of strong oscillation in
\rf{anizo_eros}. In the previous case the fast change in the boundary
condition causes stronger transients than it was by the Kelvin model,
while in the latter case the solutions show strong oscillations, as we
have expected.
 
We also present the solution of a cylindrical bore (tunnel) opened in
homogeneous medium loaded by its self weight when no lateral deformations
are allowed in the primary field
\m{(k=\f{\nu}{1-\nu}=\f{1-\qeta}{1+2\qeta})}. In case of stongly damped
oscillations the time evolution of the coefficient functions are shown
in \rf{gatolt_gyenge}, and the case of weakly damped
oscillations are plotted in \rf{gatolt_eros}. In the previous
case one can see weak oscillation in the initial transients, while in
the latter case the oscillation can be observed on each coefficient
functions.
 
Since materials with positive index of inertia are currently not known,
the corresponding displacement fields are not presented.

\fffigg{0.475}{anizotrop_alagut_kl_v_s_lassu_1_color.pdf}
              {anizotrop_alagut_kl_v_s_koz_1_color.pdf}
              {anizotrop_alagut_kl_v_s_gyors_1_color.pdf}
              {anizotrop_alagut_kl_v_z_lassu_1_color.pdf}
              {anizotrop_alagut_kl_v_z_koz_1_color.pdf}
              {anizotrop_alagut_kl_v_z_gyors_1_color.pdf}
{Weak oscillations of the solution of a cylindrical bore (tunnel) opened
in homogeneous but anisotropic stress field.
 \LeftRight{s}{s}  }{anizo_gyenge}

\fffigg{0.475}{anizotrop_alagut_kl_v_s_lassu_color.pdf}
              {anizotrop_alagut_kl_v_s_koz_color.pdf}
              {anizotrop_alagut_kl_v_s_gyors_color.pdf}
              {anizotrop_alagut_kl_v_z_lassu_color.pdf}
              {anizotrop_alagut_kl_v_z_koz_color.pdf}
              {anizotrop_alagut_kl_v_z_gyors_color.pdf}
{Strong oscillations of the solution of a cylindrical bore (tunnel)
opened in homogeneous but anisotropic stress field.
 \LeftRight{s}{s}  }{anizo_eros}

\fffigg{0.475}{onsuly_lamnu_kl_v_s_lassu_1_color.pdf}
              {onsuly_lamnu_kl_v_s_koz_1_color.pdf}
              {onsuly_lamnu_kl_v_s_gyors_1_color.pdf}
              {onsuly_lamnu_kl_v_z_lassu_1_color.pdf}
              {onsuly_lamnu_kl_v_z_koz_1_color.pdf}
              {onsuly_lamnu_kl_v_z_gyors_1_color.pdf}
{Weak oscillations of the solution of a cylindrical bore (tunnel) opened
in homogeneous medium loaded by its self weight, with no lateral
deformations allowed in the primary field
(\m{k=\f{\nu}{1-\nu}=\f{1-\qeta}{1+2\qeta}}).
 \LeftRight{s}{s}  }{gatolt_gyenge}

\fffigg{0.475}{onsuly_lamnu_kl_v_s_lassu_color.pdf}
              {onsuly_lamnu_kl_v_s_koz_color.pdf}
              {onsuly_lamnu_kl_v_s_gyors_color.pdf}
              {onsuly_lamnu_kl_v_z_lassu_color.pdf}
              {onsuly_lamnu_kl_v_z_koz_color.pdf}
              {onsuly_lamnu_kl_v_z_gyors_color.pdf}
{Strong oscillations of the solution of a cylindrical bore (tunnel)
opened in homogeneous medium loaded by its self weight, with no lateral
deformations allowed in the primary field
(\m{k=\f{\nu}{1-\nu}=\f{1-\qeta}{1+2\qeta}}).
 \LeftRight{s}{s}  }{gatolt_eros}

\subsection{Displacement fields}

The displacement field\footnote{Understood, naturally, with respect to
the primary initial state of the continuum.} can be given via the
Ces\`aro formula \re{cesaro}. One needs to substitute into \re{cesaro}
\m{\hheps=\f{1}{\qEs}\hhzet} based on \re{feszdim_al}, and then
\re{reolzeta} or \re{zetreol3}. Since stress-dimensioned strain is given
in finite sum form and, moreover, spatial and time dependences are
separated, one can obtain
 \begin{align}
 \qqu^{\rm Cauchy}\71{t,\qr} = \qqu_0\01{t} & + \qOm\81{t}\01{\qr-\qr_0}
 \mathrel+
\nonumber  \\
 &
+\f{1}{\qEs}\sum_{j=1}^J\dev{\rm{K}}_j\01{t}
\int_{\qr_0}^\qr\93{\dev{\qs}_j\01{\qrt}+2\92{\dev{\qs}_j\01{\qrt}
\otimes\nablal}^{{\rm A}_{1,3}}\01{\qr-\qrt}}\dd \qrt
\mathrel+
 \\  \nonumber
 &
+\f{1}{\qEs}\sum_{j=1}^J\sph{\rm{K}}_j\01{t}
\int_{\qr_0}^\qr\93{\sph{\qs}_j\01{\qrt}+2\92{\sph{\qs}_j\01{\qrt}
\otimes\nablal}^{{\rm A}_{1,3}}\01{\qr-\qrt}}\dd \qrt
 \end{align}
or
 \begin{align}
\qqu^{\rm Cauchy}\71{t,\qr}=\qqu_0\01{t} & +\qOm\81{t}\01{\qr-\qr_0}
\mathrel+
 \nonumber  \\
& +\f{1}{\qEs}\sum_{k=1}^K \91{ \qPsid_k\01{t}
\int_{\qr_0}^\qr\93{\qgamma_k\01{\qrt}+2\92{\qgamma_k\01{\qrt}
\otimes\nablal}^{{\rm A}_{1,3}}\01{\qr-\qrt}}\dd \qrt}
\mathrel+
 \\  \nonumber
& +\f{1}{\qEs}\sum_{l=1}^L\91{ \qPsis_l\01{t} \int_{\qr_0}^\qr
\93{\qdelta_l\01{\qrt}+2\92{\qdelta_l\01{\qrt}
\otimes\nablal}^{{\rm A}_{1,3}}\01{\qr-\qrt}}\dd \qrt} ,
 \end{align}
where
\m { \qPsid_k }, \m { \qPsis_l } are defined in \re{hehe2}.
One should not forget that these are only Cauchy
vector potentials so the rigid body like displacement and rotation have
to be fixed.
 
In case of the cylindrical bore (tunnel) opened in infinite, homogeneous
and isotropic or anisotropic stress field, these uncertainties can be
completely fixed by integrating from the centre and by subtracting
the rotation in the infinity.
 
To fix these uncertainties for a cylindrical bore (tunnel) opened in
homogeneous medium loaded by its self weight is not so easy. The elastic
solution, which is used to derive the rheological solution, is only the
first-term approximation of an infinite series so our rheological
solution is just an approximation, valid only in a certain neighbourhood
of the bore. The Cauchy vector potentials that originate from these
solutions diverge at infinity so fixing the uncertainties at infinity
cannot be used.\footnote{Rotation does not pose a problem but \m {
\qqu_0\01{t} } does.} Instead of this, at each time instant, we
determine the point of the surface where the tangent is zero, and
subtract the vertical displacement of this point from the displacement
field, which procedure turns out to provide reasonably realistic result
(see the figures).
 
In the example of a cylindrical bore (tunnel) opened in homogeneous
medium loaded by its self weight, we have chosen the dimensionless
ratios
 \begin{align}
d/R = 2,  \qquad\quad  \gamma / \qEs = 0.02
 \end{align}
for the calculated results shown here. For the cylindrical bore (tunnel)
opened in homogeneous and isotropic stress field, the used stress
component \m{\bsig_{rr}} has been adjusted correspondingly:
\m{\bsig_{rr}=-\gamma d}. Similarly, for the problem of the cylindrical
bore (tunnel) opened in homogeneous but anisotropic stress field we took
the stress component \m{\bsig_{yy}=-\gamma d}, and the stress component
\m{\bsig_{xx}} was be calculated from \m{\bsig_{yy}} via the lateral
pressure factor \m{k}, applying the value
\m{k=\f{\nu}{1-\nu}=\f{1-\qeta}{1+2\qeta}}, which describes the case
when lateral deformations are not allowed in the primary field. We chose
\m{\bsig_{xy}=-\gamma d}, and zeros for the other stress
components because we analyze the problem perpendicular to the axis of
the bore.
 
\rff{u_izo}--\ref{u_szabad} show the time evolution of the displacement
field for the cases of slow, medium and fast changes in the boundary
condition. Time goes by from the blue line towards the red line; the
contour of the bore and the surface are plotted in the nondimensional
time instants 0, 0.5, 1, 2, 4 and 8. In the figures, the displacement of
the contours has been artificially enlarged since the found small-strain
displacements themselves would be too small for the eye, and our primary
goal here is the visualisation of the \emph{tendencies}.
 
Generally, it can be stated that the shape of the bore changes weakly at
the beginning of the process for slow changes in the boundary condition,
however, when the boundary condition changes fast then the bore rapidly
deforms at the beginning of the process. According to our expectations,
the elastic final (asymptotic) state is independent of the changes of
the boundary condition.
 
\rf{u_izo} shows the time evolution of the displacement field of a
cylindrical bore (tunnel) opened in homogeneous and isotropic stress
field. The bore is shrunk as time goes by and, at the same time, the
surface is displaced in the direction of the center of the bore.
 
\rf{u_anizo} shows the time evolution of the displacement field of a
cylindrical bore (tunnel) opened in homogeneous but anisotropic stress
field. Although this model does not take into account the load of the
self weight of the medium, it is possible to calculate the rotation of
the cross section using this approximation. Other interesting
observations can also be made: we can notice that the geometry is
stretched in the direction of the major axis of the rotated ellipse and
shrinks only later. Comparing these with \rf{anizo}, we can see that
one of the coefficient functions starts with anomalous sign at the
beginning of the process so the pattern that belongs to this function
initially expands and shrinks only later.
 
\rff{u_hidro}--\ref{u_szabad} show the time evolution of the
displacement field belonging to the cylindrical bore (tunnel) opened in
homogeneous medium loaded by its self weight. This model approximates
the processes around tunnels the best. One can observe in each figure
that the center of the bore moves towards the direction of the surface.
In case of hydrostatic initial stress state (\rf{hidro}), the contour
of the bore deforms approximately isotropically, and the sinking of the
surface is negligible. When lateral deformations in the primary field
are not allowed (\rf{u_gatolt}), and in the opposite case
(\rf{u_szabad}) the sinking of the surface and a swelling-like
phenomenon are also observable.

\newpage

\fffiG{izo_lassu_reol_elmozd_fig_color.pdf}
      {izo_koz_reol_elmozd_fig_color.pdf}
      {izo_gyors_reol_elmozd_fig_color.pdf}
{Time evolution tendency of the displacement field of a cylindrical bore
(tunnel) opened in homogeneous and isotropic stress field.
 \UpDown}  {u_izo}

\fffiG{anizo_lassu_reol_elmozd_fig_color.pdf}
      {anizo_koz_reol_elmozd_fig_color.pdf}
      {anizo_gyors_reol_elmozd_fig_color.pdf}
{Time evolution tendency of the displacement field of a cylindrical bore
(tunnel) opened in homogeneous but anisotropic stress field.
 \UpDown}  {u_anizo}

\fffiG{onsuly_hidro_lassu_reol_elmozd_fig_color.pdf}
      {onsuly_hidro_koz_reol_elmozd_fig_color.pdf}
      {onsuly_hidro_gyors_reol_elmozd_fig_color.pdf}
{Time evolution tendency of the displacement field of a cylindrical bore
(tunnel) opened in homogeneous medium loaded by its self weight,
hydrostatic initial stress state (\m{k=1}).
 \UpDown}  {u_hidro}

\fffiG{onsuly_gatolt_lassu_reol_elmozd_fig_color.pdf}
      {onsuly_gatolt_koz_reol_elmozd_fig_color.pdf}
      {onsuly_gatolt_gyors_reol_elmozd_fig_color.pdf}
{Time evolution tendency of the displacement field of a cylindrical bore
(tunnel) opened in homogeneous medium loaded by its self weight, no
lateral deformations allowed in the primary field
(\m{k=\f{\nu}{1-\nu}=\f{1-\qeta}{1+2\qeta}}).
 \UpDown}  {u_gatolt}

\fffiG{onsuly_szabad_lassu_reol_elmozd_fig_color.pdf}
      {onsuly_szabad_koz_reol_elmozd_fig_color.pdf}
      {onsuly_szabad_gyors_reol_elmozd_fig_color.pdf}
{Time evolution tendency of the displacement field of a cylindrical bore
(tunnel) opened in homogeneous medium loaded by its self weight, free
lateral deformations in the primary field (\m{k=0}).
 \UpDown}  {u_szabad}

\Section{Conclusions and outlook}

The method presented here (in its most general, third, form) is
applicable to many rheological problems. It can be helpful \eg during
the process of designing of tunnels and underground facilities as an
insightful approximation, but can also be useful for validating
numerical solvers.

As a heuristic summation of the range of problems that can be covered by
this method is that problems that are simple enough to be treatable via
reasonable analytic means can probably be solved this way.

Mathematically, a limitation of the method is that initial conditions
must be expressible in terms of the utilized elastic spatial patterns.
Fortunately, the vast majority of practical problems---those when one
assumes a stationary, undisturbed, equilibrial initial state for a
time interval---is of this kind.

The development of the method can continue in the future in several
directions. First, it would be instructive and useful to compare
obtained results with ones stemming from other methods (\eg finite
element method).
 
Second, in each of the considered examples, the process was a plane-strain
process. In this case, the form of the strain tensor is
 \begin{align}
\hheps&=\begin{pmatrix}\heps_{11}&\heps_{12}&0\\\heps_{12}&\heps_{22}&0\\0&0&0\end{pmatrix}
 \end{align} 
in an appropriate coordinate system, and the corresponding elastic stress
tensor is
 \begin{align}
\hhsig&=\begin{pmatrix}\hsig_{11}&\hsig_{12}&0\\\hsig_{12}&\hsig_{22}&0\\0&0&\nu\left(\hsig_{11}+\hsig_{22}\right)\end{pmatrix}.
 \end{align}
This seems to be the reason why in the considered examples we always
happen to find a relationship among the spherical parts of the
independent spatial patterns, which reduces the number of independent
rheological equations so one arrives at a well-determined system of
linear ordinary differential equations. Our conjecture is that the
method can be applied for all plane problems. This statement has not
proven yet but in all the considered examples this property can be seen
so it is plausible to search for a proof of it.

When there is no relationship among the spherical patterns then the method
results in an overdetermined differential equation system. In this case, a
generalized, approximate, version of the method could be used which
would provide an optimized approximate solution. In this context,
variational formulation of the rheological problem may be fruitful.

Finally, the method could be generalized to cases when the elastic
stress solution can be given only in infinite sum form (see \eg
\cite{mindlin}), or the recent remarkable, tunnel related, analytical
solution by Cai and coworkers \cite{Cai2019}).
The approximation of the infinitely many terms with a
finite sum can lead to a novel numerical method that may be capable to
treat situations with complex geometries. For this purpose, a
variational formulation can also come helpful.

\subsubsection*{Acknowledgement}

The authors thank Csaba Asszonyi and Zolt\'an Szarka for inspiring
conversations and ideas, Gyula B\'eda for valuable discussions,
and L\'aszl\'o Szab\'o for useful suggestions.
The work was supported by the National Research, Development and
Innovation Office NKFIH KH 130378, the National Research, Development
and Innovation Fund (TUDFO/51757/2019-ITM), Thematic Excellence
Program and the R\&D project NO.\ 2018-1.1.2-KFI-2018-00207.

\end{document}